\documentclass[longauth]{aaEC}

\usepackage{graphicx}
\usepackage{natbib}
\usepackage{scalerel}
\usepackage{siunitx}

\usepackage[table]{xcolor}

\usepackage{txfonts}
\usepackage[pdfencoding=auto,psdextra]{hyperref}
\hypersetup{
    colorlinks=true,
    linkcolor=blue,
    filecolor=magenta,      
    urlcolor=blue,
    citecolor=blue
}
\makeatletter
\renewcommand*\aa@pageof{, page \thepage{} of \pageref*{LastPage}}
\makeatother

\usepackage[utf8]{inputenc}

\usepackage[switch, modulo]{lineno}

\usepackage{euclid}

\begin{document}
%
%

\title{\Euclid\/: Quick Data Release (Q1) -- Watching ICM-selected galaxy clusters with \Euclid eyes - prospects of \Euclid data in the context of large SZ and X-ray based surveys\thanks{This paper is published on
       behalf of the Euclid Consortium}}

   

\newcommand{\orcid}[1]{} 
\author{M.~Klein\orcid{0000-0002-8248-4488}\thanks{\email{matthias.klein@physik.uni-muenchen.de}}\inst{\ref{aff1}}
\and K.~George\orcid{0000-0002-1734-8455}\inst{\ref{aff1}}
\and J.~J.~Mohr\orcid{0000-0002-6875-2087}\inst{\ref{aff1}}
\and B.~Altieri\orcid{0000-0003-3936-0284}\inst{\ref{aff2}}
\and L.~Amendola\orcid{0000-0002-0835-233X}\inst{\ref{aff3}}
\and S.~Andreon\orcid{0000-0002-2041-8784}\inst{\ref{aff4}}
\and N.~Auricchio\orcid{0000-0003-4444-8651}\inst{\ref{aff5}}
\and C.~Baccigalupi\orcid{0000-0002-8211-1630}\inst{\ref{aff6},\ref{aff7},\ref{aff8},\ref{aff9}}
\and M.~Baldi\orcid{0000-0003-4145-1943}\inst{\ref{aff10},\ref{aff5},\ref{aff11}}
\and A.~Balestra\orcid{0000-0002-6967-261X}\inst{\ref{aff12}}
\and S.~Bardelli\orcid{0000-0002-8900-0298}\inst{\ref{aff5}}
\and A.~Biviano\orcid{0000-0002-0857-0732}\inst{\ref{aff7},\ref{aff6}}
\and E.~Branchini\orcid{0000-0002-0808-6908}\inst{\ref{aff13},\ref{aff14},\ref{aff4}}
\and M.~Brescia\orcid{0000-0001-9506-5680}\inst{\ref{aff15},\ref{aff16}}
\and S.~Camera\orcid{0000-0003-3399-3574}\inst{\ref{aff17},\ref{aff18},\ref{aff19}}
\and G.~Ca\~nas-Herrera\orcid{0000-0003-2796-2149}\inst{\ref{aff20},\ref{aff21},\ref{aff22}}
\and V.~Capobianco\orcid{0000-0002-3309-7692}\inst{\ref{aff19}}
\and C.~Carbone\orcid{0000-0003-0125-3563}\inst{\ref{aff23}}
\and J.~Carretero\orcid{0000-0002-3130-0204}\inst{\ref{aff24},\ref{aff25}}
\and S.~Casas\orcid{0000-0002-4751-5138}\inst{\ref{aff26}}
\and M.~Castellano\orcid{0000-0001-9875-8263}\inst{\ref{aff27}}
\and G.~Castignani\orcid{0000-0001-6831-0687}\inst{\ref{aff5}}
\and S.~Cavuoti\orcid{0000-0002-3787-4196}\inst{\ref{aff16},\ref{aff28}}
\and K.~C.~Chambers\orcid{0000-0001-6965-7789}\inst{\ref{aff29}}
\and A.~Cimatti\inst{\ref{aff30}}
\and C.~Colodro-Conde\inst{\ref{aff31}}
\and G.~Congedo\orcid{0000-0003-2508-0046}\inst{\ref{aff32}}
\and L.~Conversi\orcid{0000-0002-6710-8476}\inst{\ref{aff33},\ref{aff2}}
\and Y.~Copin\orcid{0000-0002-5317-7518}\inst{\ref{aff34}}
\and F.~Courbin\orcid{0000-0003-0758-6510}\inst{\ref{aff35},\ref{aff36}}
\and H.~M.~Courtois\orcid{0000-0003-0509-1776}\inst{\ref{aff37}}
\and M.~Cropper\orcid{0000-0003-4571-9468}\inst{\ref{aff38}}
\and A.~Da~Silva\orcid{0000-0002-6385-1609}\inst{\ref{aff39},\ref{aff40}}
\and H.~Degaudenzi\orcid{0000-0002-5887-6799}\inst{\ref{aff41}}
\and G.~De~Lucia\orcid{0000-0002-6220-9104}\inst{\ref{aff7}}
\and C.~Dolding\orcid{0009-0003-7199-6108}\inst{\ref{aff38}}
\and H.~Dole\orcid{0000-0002-9767-3839}\inst{\ref{aff42}}
\and F.~Dubath\orcid{0000-0002-6533-2810}\inst{\ref{aff41}}
\and F.~Ducret\inst{\ref{aff43}}
\and X.~Dupac\inst{\ref{aff2}}
\and S.~Dusini\orcid{0000-0002-1128-0664}\inst{\ref{aff44}}
\and M.~Farina\orcid{0000-0002-3089-7846}\inst{\ref{aff45}}
\and R.~Farinelli\inst{\ref{aff5}}
\and F.~Faustini\orcid{0000-0001-6274-5145}\inst{\ref{aff27},\ref{aff46}}
\and S.~Ferriol\inst{\ref{aff34}}
\and F.~Finelli\orcid{0000-0002-6694-3269}\inst{\ref{aff5},\ref{aff47}}
\and M.~Frailis\orcid{0000-0002-7400-2135}\inst{\ref{aff7}}
\and E.~Franceschi\orcid{0000-0002-0585-6591}\inst{\ref{aff5}}
\and M.~Fumana\orcid{0000-0001-6787-5950}\inst{\ref{aff23}}
\and S.~Galeotta\orcid{0000-0002-3748-5115}\inst{\ref{aff7}}
\and B.~Gillis\orcid{0000-0002-4478-1270}\inst{\ref{aff32}}
\and C.~Giocoli\orcid{0000-0002-9590-7961}\inst{\ref{aff5},\ref{aff11}}
\and J.~Gracia-Carpio\inst{\ref{aff48}}
\and A.~Grazian\orcid{0000-0002-5688-0663}\inst{\ref{aff12}}
\and F.~Grupp\inst{\ref{aff48},\ref{aff49}}
\and S.~V.~H.~Haugan\orcid{0000-0001-9648-7260}\inst{\ref{aff50}}
\and W.~Holmes\inst{\ref{aff51}}
\and I.~M.~Hook\orcid{0000-0002-2960-978X}\inst{\ref{aff52}}
\and F.~Hormuth\inst{\ref{aff53}}
\and A.~Hornstrup\orcid{0000-0002-3363-0936}\inst{\ref{aff54},\ref{aff55}}
\and K.~Jahnke\orcid{0000-0003-3804-2137}\inst{\ref{aff56}}
\and M.~Jhabvala\inst{\ref{aff57}}
\and E.~Keih\"anen\orcid{0000-0003-1804-7715}\inst{\ref{aff58}}
\and S.~Kermiche\orcid{0000-0002-0302-5735}\inst{\ref{aff59}}
\and B.~Kubik\orcid{0009-0006-5823-4880}\inst{\ref{aff34}}
\and M.~K\"ummel\orcid{0000-0003-2791-2117}\inst{\ref{aff1}}
\and M.~Kunz\orcid{0000-0002-3052-7394}\inst{\ref{aff60}}
\and H.~Kurki-Suonio\orcid{0000-0002-4618-3063}\inst{\ref{aff61},\ref{aff62}}
\and A.~M.~C.~Le~Brun\orcid{0000-0002-0936-4594}\inst{\ref{aff63}}
\and D.~Le~Mignant\orcid{0000-0002-5339-5515}\inst{\ref{aff43}}
\and S.~Ligori\orcid{0000-0003-4172-4606}\inst{\ref{aff19}}
\and P.~B.~Lilje\orcid{0000-0003-4324-7794}\inst{\ref{aff50}}
\and V.~Lindholm\orcid{0000-0003-2317-5471}\inst{\ref{aff61},\ref{aff62}}
\and I.~Lloro\orcid{0000-0001-5966-1434}\inst{\ref{aff64}}
\and G.~Mainetti\orcid{0000-0003-2384-2377}\inst{\ref{aff65}}
\and D.~Maino\inst{\ref{aff66},\ref{aff23},\ref{aff67}}
\and E.~Maiorano\orcid{0000-0003-2593-4355}\inst{\ref{aff5}}
\and O.~Mansutti\orcid{0000-0001-5758-4658}\inst{\ref{aff7}}
\and O.~Marggraf\orcid{0000-0001-7242-3852}\inst{\ref{aff68}}
\and M.~Martinelli\orcid{0000-0002-6943-7732}\inst{\ref{aff27},\ref{aff69}}
\and N.~Martinet\orcid{0000-0003-2786-7790}\inst{\ref{aff43}}
\and F.~Marulli\orcid{0000-0002-8850-0303}\inst{\ref{aff70},\ref{aff5},\ref{aff11}}
\and R.~Massey\orcid{0000-0002-6085-3780}\inst{\ref{aff71}}
\and S.~Maurogordato\inst{\ref{aff72}}
\and E.~Medinaceli\orcid{0000-0002-4040-7783}\inst{\ref{aff5}}
\and S.~Mei\orcid{0000-0002-2849-559X}\inst{\ref{aff73},\ref{aff74}}
\and Y.~Mellier\inst{\ref{aff75},\ref{aff76}}
\and M.~Meneghetti\orcid{0000-0003-1225-7084}\inst{\ref{aff5},\ref{aff11}}
\and E.~Merlin\orcid{0000-0001-6870-8900}\inst{\ref{aff27}}
\and G.~Meylan\inst{\ref{aff77}}
\and L.~Moscardini\orcid{0000-0002-3473-6716}\inst{\ref{aff70},\ref{aff5},\ref{aff11}}
\and R.~Nakajima\orcid{0009-0009-1213-7040}\inst{\ref{aff68}}
\and C.~Neissner\orcid{0000-0001-8524-4968}\inst{\ref{aff78},\ref{aff25}}
\and S.-M.~Niemi\orcid{0009-0005-0247-0086}\inst{\ref{aff20}}
\and C.~Padilla\orcid{0000-0001-7951-0166}\inst{\ref{aff78}}
\and S.~Paltani\orcid{0000-0002-8108-9179}\inst{\ref{aff41}}
\and F.~Pasian\orcid{0000-0002-4869-3227}\inst{\ref{aff7}}
\and K.~Pedersen\inst{\ref{aff79}}
\and W.~J.~Percival\orcid{0000-0002-0644-5727}\inst{\ref{aff80},\ref{aff81},\ref{aff82}}
\and V.~Pettorino\inst{\ref{aff20}}
\and S.~Pires\orcid{0000-0002-0249-2104}\inst{\ref{aff83}}
\and G.~Polenta\orcid{0000-0003-4067-9196}\inst{\ref{aff46}}
\and M.~Poncet\inst{\ref{aff84}}
\and L.~A.~Popa\inst{\ref{aff85}}
\and L.~Pozzetti\orcid{0000-0001-7085-0412}\inst{\ref{aff5}}
\and F.~Raison\orcid{0000-0002-7819-6918}\inst{\ref{aff48}}
\and R.~Rebolo\orcid{0000-0003-3767-7085}\inst{\ref{aff31},\ref{aff86},\ref{aff87}}
\and A.~Renzi\orcid{0000-0001-9856-1970}\inst{\ref{aff88},\ref{aff44}}
\and J.~Rhodes\orcid{0000-0002-4485-8549}\inst{\ref{aff51}}
\and G.~Riccio\inst{\ref{aff16}}
\and E.~Romelli\orcid{0000-0003-3069-9222}\inst{\ref{aff7}}
\and M.~Roncarelli\orcid{0000-0001-9587-7822}\inst{\ref{aff5}}
\and C.~Rosset\orcid{0000-0003-0286-2192}\inst{\ref{aff73}}
\and H.~J.~A.~Rottgering\orcid{0000-0001-8887-2257}\inst{\ref{aff22}}
\and R.~Saglia\orcid{0000-0003-0378-7032}\inst{\ref{aff49},\ref{aff48}}
\and Z.~Sakr\orcid{0000-0002-4823-3757}\inst{\ref{aff3},\ref{aff89},\ref{aff90}}
\and D.~Sapone\orcid{0000-0001-7089-4503}\inst{\ref{aff91}}
\and M.~Schirmer\orcid{0000-0003-2568-9994}\inst{\ref{aff56}}
\and P.~Schneider\orcid{0000-0001-8561-2679}\inst{\ref{aff68}}
\and T.~Schrabback\orcid{0000-0002-6987-7834}\inst{\ref{aff92}}
\and A.~Secroun\orcid{0000-0003-0505-3710}\inst{\ref{aff59}}
\and E.~Sefusatti\orcid{0000-0003-0473-1567}\inst{\ref{aff7},\ref{aff6},\ref{aff8}}
\and G.~Seidel\orcid{0000-0003-2907-353X}\inst{\ref{aff56}}
\and S.~Serrano\orcid{0000-0002-0211-2861}\inst{\ref{aff93},\ref{aff94},\ref{aff95}}
\and C.~Sirignano\orcid{0000-0002-0995-7146}\inst{\ref{aff88},\ref{aff44}}
\and G.~Sirri\orcid{0000-0003-2626-2853}\inst{\ref{aff11}}
\and L.~Stanco\orcid{0000-0002-9706-5104}\inst{\ref{aff44}}
\and J.~Steinwagner\orcid{0000-0001-7443-1047}\inst{\ref{aff48}}
\and P.~Tallada-Cresp\'{i}\orcid{0000-0002-1336-8328}\inst{\ref{aff24},\ref{aff25}}
\and A.~N.~Taylor\inst{\ref{aff32}}
\and I.~Tereno\orcid{0000-0002-4537-6218}\inst{\ref{aff39},\ref{aff96}}
\and S.~Toft\orcid{0000-0003-3631-7176}\inst{\ref{aff97},\ref{aff98}}
\and R.~Toledo-Moreo\orcid{0000-0002-2997-4859}\inst{\ref{aff99}}
\and F.~Torradeflot\orcid{0000-0003-1160-1517}\inst{\ref{aff25},\ref{aff24}}
\and I.~Tutusaus\orcid{0000-0002-3199-0399}\inst{\ref{aff89}}
\and L.~Valenziano\orcid{0000-0002-1170-0104}\inst{\ref{aff5},\ref{aff47}}
\and J.~Valiviita\orcid{0000-0001-6225-3693}\inst{\ref{aff61},\ref{aff62}}
\and T.~Vassallo\orcid{0000-0001-6512-6358}\inst{\ref{aff1},\ref{aff7}}
\and A.~Veropalumbo\orcid{0000-0003-2387-1194}\inst{\ref{aff4},\ref{aff14},\ref{aff13}}
\and Y.~Wang\orcid{0000-0002-4749-2984}\inst{\ref{aff100}}
\and J.~Weller\orcid{0000-0002-8282-2010}\inst{\ref{aff49},\ref{aff48}}
\and F.~M.~Zerbi\inst{\ref{aff4}}
\and E.~Zucca\orcid{0000-0002-5845-8132}\inst{\ref{aff5}}
\and C.~Burigana\orcid{0000-0002-3005-5796}\inst{\ref{aff101},\ref{aff47}}
\and V.~Scottez\orcid{0009-0008-3864-940X}\inst{\ref{aff75},\ref{aff102}}
\and M.~Sereno\orcid{0000-0003-0302-0325}\inst{\ref{aff5},\ref{aff11}}
\and M.~Viel\orcid{0000-0002-2642-5707}\inst{\ref{aff6},\ref{aff7},\ref{aff9},\ref{aff8},\ref{aff103}}}
										   
\institute{University Observatory, LMU Faculty of Physics, Scheinerstrasse 1, 81679 Munich, Germany\label{aff1}
\and
ESAC/ESA, Camino Bajo del Castillo, s/n., Urb. Villafranca del Castillo, 28692 Villanueva de la Ca\~nada, Madrid, Spain\label{aff2}
\and
Institut f\"ur Theoretische Physik, University of Heidelberg, Philosophenweg 16, 69120 Heidelberg, Germany\label{aff3}
\and
INAF-Osservatorio Astronomico di Brera, Via Brera 28, 20122 Milano, Italy\label{aff4}
\and
INAF-Osservatorio di Astrofisica e Scienza dello Spazio di Bologna, Via Piero Gobetti 93/3, 40129 Bologna, Italy\label{aff5}
\and
IFPU, Institute for Fundamental Physics of the Universe, via Beirut 2, 34151 Trieste, Italy\label{aff6}
\and
INAF-Osservatorio Astronomico di Trieste, Via G. B. Tiepolo 11, 34143 Trieste, Italy\label{aff7}
\and
INFN, Sezione di Trieste, Via Valerio 2, 34127 Trieste TS, Italy\label{aff8}
\and
SISSA, International School for Advanced Studies, Via Bonomea 265, 34136 Trieste TS, Italy\label{aff9}
\and
Dipartimento di Fisica e Astronomia, Universit\`a di Bologna, Via Gobetti 93/2, 40129 Bologna, Italy\label{aff10}
\and
INFN-Sezione di Bologna, Viale Berti Pichat 6/2, 40127 Bologna, Italy\label{aff11}
\and
INAF-Osservatorio Astronomico di Padova, Via dell'Osservatorio 5, 35122 Padova, Italy\label{aff12}
\and
Dipartimento di Fisica, Universit\`a di Genova, Via Dodecaneso 33, 16146, Genova, Italy\label{aff13}
\and
INFN-Sezione di Genova, Via Dodecaneso 33, 16146, Genova, Italy\label{aff14}
\and
Department of Physics "E. Pancini", University Federico II, Via Cinthia 6, 80126, Napoli, Italy\label{aff15}
\and
INAF-Osservatorio Astronomico di Capodimonte, Via Moiariello 16, 80131 Napoli, Italy\label{aff16}
\and
Dipartimento di Fisica, Universit\`a degli Studi di Torino, Via P. Giuria 1, 10125 Torino, Italy\label{aff17}
\and
INFN-Sezione di Torino, Via P. Giuria 1, 10125 Torino, Italy\label{aff18}
\and
INAF-Osservatorio Astrofisico di Torino, Via Osservatorio 20, 10025 Pino Torinese (TO), Italy\label{aff19}
\and
European Space Agency/ESTEC, Keplerlaan 1, 2201 AZ Noordwijk, The Netherlands\label{aff20}
\and
Institute Lorentz, Leiden University, Niels Bohrweg 2, 2333 CA Leiden, The Netherlands\label{aff21}
\and
Leiden Observatory, Leiden University, Einsteinweg 55, 2333 CC Leiden, The Netherlands\label{aff22}
\and
INAF-IASF Milano, Via Alfonso Corti 12, 20133 Milano, Italy\label{aff23}
\and
Centro de Investigaciones Energ\'eticas, Medioambientales y Tecnol\'ogicas (CIEMAT), Avenida Complutense 40, 28040 Madrid, Spain\label{aff24}
\and
Port d'Informaci\'{o} Cient\'{i}fica, Campus UAB, C. Albareda s/n, 08193 Bellaterra (Barcelona), Spain\label{aff25}
\and
Institute for Theoretical Particle Physics and Cosmology (TTK), RWTH Aachen University, 52056 Aachen, Germany\label{aff26}
\and
INAF-Osservatorio Astronomico di Roma, Via Frascati 33, 00078 Monteporzio Catone, Italy\label{aff27}
\and
INFN section of Naples, Via Cinthia 6, 80126, Napoli, Italy\label{aff28}
\and
Institute for Astronomy, University of Hawaii, 2680 Woodlawn Drive, Honolulu, HI 96822, USA\label{aff29}
\and
Dipartimento di Fisica e Astronomia "Augusto Righi" - Alma Mater Studiorum Universit\`a di Bologna, Viale Berti Pichat 6/2, 40127 Bologna, Italy\label{aff30}
\and
Instituto de Astrof\'{\i}sica de Canarias, V\'{\i}a L\'actea, 38205 La Laguna, Tenerife, Spain\label{aff31}
\and
Institute for Astronomy, University of Edinburgh, Royal Observatory, Blackford Hill, Edinburgh EH9 3HJ, UK\label{aff32}
\and
European Space Agency/ESRIN, Largo Galileo Galilei 1, 00044 Frascati, Roma, Italy\label{aff33}
\and
Universit\'e Claude Bernard Lyon 1, CNRS/IN2P3, IP2I Lyon, UMR 5822, Villeurbanne, F-69100, France\label{aff34}
\and
Institut de Ci\`{e}ncies del Cosmos (ICCUB), Universitat de Barcelona (IEEC-UB), Mart\'{i} i Franqu\`{e}s 1, 08028 Barcelona, Spain\label{aff35}
\and
Instituci\'o Catalana de Recerca i Estudis Avan\c{c}ats (ICREA), Passeig de Llu\'{\i}s Companys 23, 08010 Barcelona, Spain\label{aff36}
\and
UCB Lyon 1, CNRS/IN2P3, IUF, IP2I Lyon, 4 rue Enrico Fermi, 69622 Villeurbanne, France\label{aff37}
\and
Mullard Space Science Laboratory, University College London, Holmbury St Mary, Dorking, Surrey RH5 6NT, UK\label{aff38}
\and
Departamento de F\'isica, Faculdade de Ci\^encias, Universidade de Lisboa, Edif\'icio C8, Campo Grande, PT1749-016 Lisboa, Portugal\label{aff39}
\and
Instituto de Astrof\'isica e Ci\^encias do Espa\c{c}o, Faculdade de Ci\^encias, Universidade de Lisboa, Campo Grande, 1749-016 Lisboa, Portugal\label{aff40}
\and
Department of Astronomy, University of Geneva, ch. d'Ecogia 16, 1290 Versoix, Switzerland\label{aff41}
\and
Universit\'e Paris-Saclay, CNRS, Institut d'astrophysique spatiale, 91405, Orsay, France\label{aff42}
\and
Aix-Marseille Universit\'e, CNRS, CNES, LAM, Marseille, France\label{aff43}
\and
INFN-Padova, Via Marzolo 8, 35131 Padova, Italy\label{aff44}
\and
INAF-Istituto di Astrofisica e Planetologia Spaziali, via del Fosso del Cavaliere, 100, 00100 Roma, Italy\label{aff45}
\and
Space Science Data Center, Italian Space Agency, via del Politecnico snc, 00133 Roma, Italy\label{aff46}
\and
INFN-Bologna, Via Irnerio 46, 40126 Bologna, Italy\label{aff47}
\and
Max Planck Institute for Extraterrestrial Physics, Giessenbachstr. 1, 85748 Garching, Germany\label{aff48}
\and
Universit\"ats-Sternwarte M\"unchen, Fakult\"at f\"ur Physik, Ludwig-Maximilians-Universit\"at M\"unchen, Scheinerstrasse 1, 81679 M\"unchen, Germany\label{aff49}
\and
Institute of Theoretical Astrophysics, University of Oslo, P.O. Box 1029 Blindern, 0315 Oslo, Norway\label{aff50}
\and
Jet Propulsion Laboratory, California Institute of Technology, 4800 Oak Grove Drive, Pasadena, CA, 91109, USA\label{aff51}
\and
Department of Physics, Lancaster University, Lancaster, LA1 4YB, UK\label{aff52}
\and
Felix Hormuth Engineering, Goethestr. 17, 69181 Leimen, Germany\label{aff53}
\and
Technical University of Denmark, Elektrovej 327, 2800 Kgs. Lyngby, Denmark\label{aff54}
\and
Cosmic Dawn Center (DAWN), Denmark\label{aff55}
\and
Max-Planck-Institut f\"ur Astronomie, K\"onigstuhl 17, 69117 Heidelberg, Germany\label{aff56}
\and
NASA Goddard Space Flight Center, Greenbelt, MD 20771, USA\label{aff57}
\and
Department of Physics and Helsinki Institute of Physics, Gustaf H\"allstr\"omin katu 2, 00014 University of Helsinki, Finland\label{aff58}
\and
Aix-Marseille Universit\'e, CNRS/IN2P3, CPPM, Marseille, France\label{aff59}
\and
Universit\'e de Gen\`eve, D\'epartement de Physique Th\'eorique and Centre for Astroparticle Physics, 24 quai Ernest-Ansermet, CH-1211 Gen\`eve 4, Switzerland\label{aff60}
\and
Department of Physics, P.O. Box 64, 00014 University of Helsinki, Finland\label{aff61}
\and
Helsinki Institute of Physics, Gustaf H{\"a}llstr{\"o}min katu 2, University of Helsinki, Helsinki, Finland\label{aff62}
\and
Laboratoire d'etude de l'Univers et des phenomenes eXtremes, Observatoire de Paris, Universit\'e PSL, Sorbonne Universit\'e, CNRS, 92190 Meudon, France\label{aff63}
\and
SKA Observatory, Jodrell Bank, Lower Withington, Macclesfield, Cheshire SK11 9FT, UK\label{aff64}
\and
Centre de Calcul de l'IN2P3/CNRS, 21 avenue Pierre de Coubertin 69627 Villeurbanne Cedex, France\label{aff65}
\and
Dipartimento di Fisica "Aldo Pontremoli", Universit\`a degli Studi di Milano, Via Celoria 16, 20133 Milano, Italy\label{aff66}
\and
INFN-Sezione di Milano, Via Celoria 16, 20133 Milano, Italy\label{aff67}
\and
Universit\"at Bonn, Argelander-Institut f\"ur Astronomie, Auf dem H\"ugel 71, 53121 Bonn, Germany\label{aff68}
\and
INFN-Sezione di Roma, Piazzale Aldo Moro, 2 - c/o Dipartimento di Fisica, Edificio G. Marconi, 00185 Roma, Italy\label{aff69}
\and
Dipartimento di Fisica e Astronomia "Augusto Righi" - Alma Mater Studiorum Universit\`a di Bologna, via Piero Gobetti 93/2, 40129 Bologna, Italy\label{aff70}
\and
Department of Physics, Institute for Computational Cosmology, Durham University, South Road, Durham, DH1 3LE, UK\label{aff71}
\and
Universit\'e C\^{o}te d'Azur, Observatoire de la C\^{o}te d'Azur, CNRS, Laboratoire Lagrange, Bd de l'Observatoire, CS 34229, 06304 Nice cedex 4, France\label{aff72}
\and
Universit\'e Paris Cit\'e, CNRS, Astroparticule et Cosmologie, 75013 Paris, France\label{aff73}
\and
CNRS-UCB International Research Laboratory, Centre Pierre Bin\'etruy, IRL2007, CPB-IN2P3, Berkeley, USA\label{aff74}
\and
Institut d'Astrophysique de Paris, 98bis Boulevard Arago, 75014, Paris, France\label{aff75}
\and
Institut d'Astrophysique de Paris, UMR 7095, CNRS, and Sorbonne Universit\'e, 98 bis boulevard Arago, 75014 Paris, France\label{aff76}
\and
Institute of Physics, Laboratory of Astrophysics, Ecole Polytechnique F\'ed\'erale de Lausanne (EPFL), Observatoire de Sauverny, 1290 Versoix, Switzerland\label{aff77}
\and
Institut de F\'{i}sica d'Altes Energies (IFAE), The Barcelona Institute of Science and Technology, Campus UAB, 08193 Bellaterra (Barcelona), Spain\label{aff78}
\and
DARK, Niels Bohr Institute, University of Copenhagen, Jagtvej 155, 2200 Copenhagen, Denmark\label{aff79}
\and
Waterloo Centre for Astrophysics, University of Waterloo, Waterloo, Ontario N2L 3G1, Canada\label{aff80}
\and
Department of Physics and Astronomy, University of Waterloo, Waterloo, Ontario N2L 3G1, Canada\label{aff81}
\and
Perimeter Institute for Theoretical Physics, Waterloo, Ontario N2L 2Y5, Canada\label{aff82}
\and
Universit\'e Paris-Saclay, Universit\'e Paris Cit\'e, CEA, CNRS, AIM, 91191, Gif-sur-Yvette, France\label{aff83}
\and
Centre National d'Etudes Spatiales -- Centre spatial de Toulouse, 18 avenue Edouard Belin, 31401 Toulouse Cedex 9, France\label{aff84}
\and
Institute of Space Science, Str. Atomistilor, nr. 409 M\u{a}gurele, Ilfov, 077125, Romania\label{aff85}
\and
Consejo Superior de Investigaciones Cientificas, Calle Serrano 117, 28006 Madrid, Spain\label{aff86}
\and
Universidad de La Laguna, Departamento de Astrof\'{\i}sica, 38206 La Laguna, Tenerife, Spain\label{aff87}
\and
Dipartimento di Fisica e Astronomia "G. Galilei", Universit\`a di Padova, Via Marzolo 8, 35131 Padova, Italy\label{aff88}
\and
Institut de Recherche en Astrophysique et Plan\'etologie (IRAP), Universit\'e de Toulouse, CNRS, UPS, CNES, 14 Av. Edouard Belin, 31400 Toulouse, France\label{aff89}
\and
Universit\'e St Joseph; Faculty of Sciences, Beirut, Lebanon\label{aff90}
\and
Departamento de F\'isica, FCFM, Universidad de Chile, Blanco Encalada 2008, Santiago, Chile\label{aff91}
\and
Universit\"at Innsbruck, Institut f\"ur Astro- und Teilchenphysik, Technikerstr. 25/8, 6020 Innsbruck, Austria\label{aff92}
\and
Institut d'Estudis Espacials de Catalunya (IEEC),  Edifici RDIT, Campus UPC, 08860 Castelldefels, Barcelona, Spain\label{aff93}
\and
Satlantis, University Science Park, Sede Bld 48940, Leioa-Bilbao, Spain\label{aff94}
\and
Institute of Space Sciences (ICE, CSIC), Campus UAB, Carrer de Can Magrans, s/n, 08193 Barcelona, Spain\label{aff95}
\and
Instituto de Astrof\'isica e Ci\^encias do Espa\c{c}o, Faculdade de Ci\^encias, Universidade de Lisboa, Tapada da Ajuda, 1349-018 Lisboa, Portugal\label{aff96}
\and
Cosmic Dawn Center (DAWN)\label{aff97}
\and
Niels Bohr Institute, University of Copenhagen, Jagtvej 128, 2200 Copenhagen, Denmark\label{aff98}
\and
Universidad Polit\'ecnica de Cartagena, Departamento de Electr\'onica y Tecnolog\'ia de Computadoras,  Plaza del Hospital 1, 30202 Cartagena, Spain\label{aff99}
\and
Infrared Processing and Analysis Center, California Institute of Technology, Pasadena, CA 91125, USA\label{aff100}
\and
INAF, Istituto di Radioastronomia, Via Piero Gobetti 101, 40129 Bologna, Italy\label{aff101}
\and
ICL, Junia, Universit\'e Catholique de Lille, LITL, 59000 Lille, France\label{aff102}
\and
ICSC - Centro Nazionale di Ricerca in High Performance Computing, Big Data e Quantum Computing, Via Magnanelli 2, Bologna, Italy\label{aff103}}

%
%

\abstract{
Galaxy clusters detected through their X-ray emission or Sunyaev--Zeldovich effect (SZE), both produced by the intra-cluster medium (ICM), have been successfully used in cosmological and astrophysical studies. To maximise the scientific return and robustness of such studies, these surveys require complementary information from other datasets.
Systematic cluster confirmation and redshifts of ICM-selected cluster candidates are typically provided by wide-field optical and infrared imaging surveys, which are becoming increasingly challenged by ongoing ICM-selected samples. Particularly at high redshifts ($z>1$) reached by future SZE-selected samples, current large surveys may not be sufficient for this task. 
Deep, high-resolution infrared surveys, such as those conducted with \Euclid, are therefore essential for confirming the majority of high-redshift clusters in these future samples.
In this context, we present an analysis of the first sizeable \Euclid\ dataset (Q1), which overlaps with several ICM-selected cluster samples. We apply an adaptation of the MCMF cluster redshift and confirmation tool to \Euclid\ data to estimate key cluster properties, including redshift and richness and to predict its capabilities to confirm high-redshift galaxy clusters.
We find promising performance in redshift and richness estimation, particularly at high redshift. The performance in richness estimation at low redshifts ($z<0.4$) is currently impacted by limitations of the Q1 dataset and are likely to improve in future data releases.
By comparing MCMF measurements along random lines of sight with similar measurements from the SZE-based ACT-DR5 MCMF catalogue, we predict that the ability to confirm clusters at $1<z<2$ using \Euclid\ will be comparable to that of current large optical surveys at $z<0.6$ and will significantly enhance the capability of cluster confirmation at high redshifts. SZE-selected cluster samples will therefore especially benefit from overlap with \Euclid datasets.
Studying the five known high-$z$ SZE-selected clusters in Q1, we identify the highest-redshift jellyfish galaxy candidate found to date in an ICM-selected cluster. This galaxy, EUCL\,J035330.86$-$504347.6, is located in the massive cluster SPT-CL\,J0353$-$5043 at $z=1.32$. We also find two massive star-forming galaxies projected close to the core of ACT-CL\,J0350.0$-$4819 ($z\simeq1.46$), and evidence of strong lensing features in SPT-CL\,J0353$-$5043 and SPT-CL\,J0421$-$4845.
}

%
%
%
%
   \titlerunning{ICM-selected galaxy cluster confirmation with \Euclid}
   \authorrunning{Klein et al. }
   
   \maketitle
%
%
%
%
   
\section{\label{sc:Intro}Introduction}
\begin{figure*}
\begin{center}
\includegraphics[width=1\linewidth]{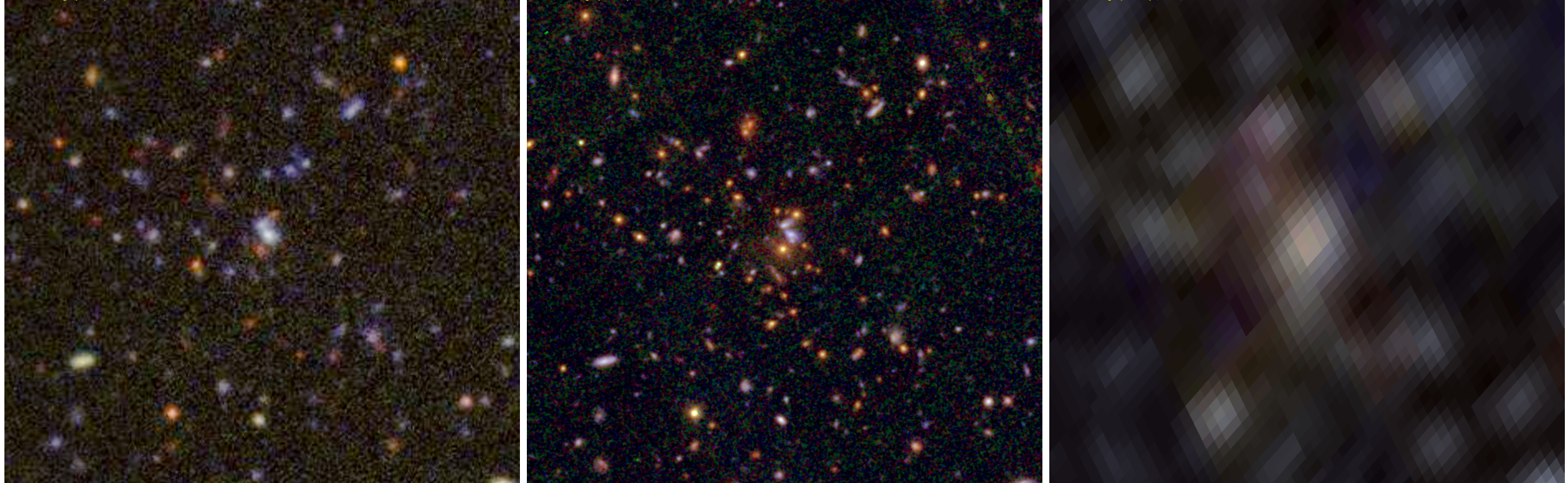}
\caption{Colour composite images of a $\ang{;1.5;}\times\ang{;1.5;}$ region centred on ACT-CL\,J0350.0$-$4819 ($z=1.46$). DESI Legacy Survey DR10, $g,r,z$ colour mage (left) used together with unWISE $w1,w2$ imaging data (right) to confirm the cluster and redshift in ACT-DR5 MCMF, compared to \Euclid \IE, \YE, \HE imaging (centre). }
\label{fig:acthighzd}
\end{center}
\end{figure*}

Galaxy cluster samples constructed through X-ray observations or through the Sunyaev-Zeldovich effect \citep[SZE,][]{Sunyaev72} signal constitute a special subclass of cluster samples as their detection observables are based on the properties of the intra-cluster medium (ICM). They have successfully been used to derive constraints on cosmological parameters, such as $\sigma_8$, $\Omega_\mathrm{m}$, and dark energy equation of state parameter $w$ \citep{WtG,2023MNRAS.522.1601C,Bocquet24b,eRASS1cosmo}. ICM-selected cluster samples were further used to study non-standard cosmology \citep{2024arXiv240913556V,mazoun2024interactingdarksectorethos,Artis24} and to constrain the Hubble parameter \citep{KozmanyanHubble,GonzalezHubble}.

The sizes of galaxy cluster samples constructed through their X-ray or SZE signature have increased by more than a factor of ten over the past decade, ranging from a few hundreds of confirmed clusters \citep{Bleem15,Bohringer:Reflex2,PSZ1cat} to now several thousands of ICM-selected clusters \citep{CODEX19,RASSMCMF,ERASS1,ACTDR5Hilton,klein2024actdr5}.
The availability of high-quality optical and infrared survey data such as from the Dark Energy Survey \citep[DES,][]{DES2016}, Sloan Digital Sky Survey \citep[SDSS,][]{SDSS}, and WISE \citep{Wright_2010} has further enhanced the characterisation of these clusters. Recent efforts have combined ICM-based selection with systematic optical follow-up using multi-band surveys \citep{MARDY3,CODEX19,2020ApJS..247...25B}, improving sample purity and completeness while extending cluster detection to lower masses and higher redshifts. The ongoing and future generation of SZE-based surveys such as SPT-3G \citep{SPT3G}, Simons Observatory \citep[SO,][]{SObs}, and CMB-S4 \citep{CMBS4white} will significantly expand current ICM-based cluster samples in the high-redshift regime, significantly challenging current ground- and space-based optical and infrared surveys. Although the upcoming Legacy Survey of Space and Time \citep[LSST,][]{LSSToverview} conducted at the Vera C. Rubin Observatory will significantly add depth compared to currently available datasets, its lack of infrared data will not allow precise redshift estimates for high-redshift galaxy clusters expected for those SZE-based surveys. In this context, the \Euclid mission \citep{EucOverview} is of key importance with its unique combination of survey area, imaging quality, and depth in the infrared regime. \Euclid\ mission itself is expected to provide $\sim10^5$ galaxy clusters selected from its optical and infrared imaging \citep{Adam-EP3}, with the potential to provide stringent cosmological constraints \citep{satoris16} if systematics on selection function and mass calibration are kept under control. First detections from the \Euclid\ galaxy cluster workflow using the Q1 dataset are presented in \cite{Q1-SP050}.

To investigate the prospects of \Euclid data to estimate redshifts and confirm ICM-selected clusters, we use the multi-component matched filter (MCMF) method \citep{2018MNRAS.474.3324K,MARDY3}, which has been successfully applied to numerous catalogues of cluster candidates from various X-ray and SZE missions and telescopes such as ROSAT \citep{Truemper93}, SPT \citep{SPT}, ACT \citep{ACTtel}, \textit{Planck} \citep{PlanckMission}, and eROSITA \citep{eROSITA}.
In this study, we apply the MCMF method on existing ICM-selected cluster catalogues RASS-MCMF \citep{RASSMCMF}, SPT-SZ MCMF \citep{SPTSZMCMF}, ACT-DR5 MCMF \citep{klein2024actdr5}, and eROSITA eRASS1 \citep{ERASS1} with the goal of evaluating and leveraging new optical and near-infrared data from the \Euclid survey. {\it Euclid\/}’s unprecedented imaging depth and homogeneous coverage provide an opportunity to improve photometric redshift estimates and identify high-redshift clusters that were previously inaccessible with ground-based optical surveys. This analysis aims to evaluate the advantages of \Euclid data in improving cluster characterisation, particularly at high redshifts, and to assess the potential systematics affecting data quality. To do so and in contrast to previous work on creating MCMF-confirmed cluster catalogues \citep[e.g. ][]{MARDY3,eFEDSKlein,2023MNRAS.525...24H}, we run MCMF on already confirmed systems and evaluate key observables such as cluster richness ($\lambda$) and cluster photometric redshifts ($z$) of those systems. 
The quality of the richness measurements, combined with MCMF runs along random lines of sight, will enable an assessment of {\it Euclid\/}’s ability to confirm high-redshift ICM-selected clusters.
As the main focus of this work is on the high-redshift ($z>1$) clusters of these samples, we further investigate and discuss the imaging and photometric data of the known high-redshift clusters in the ICM-selected samples to highlight the \Euclid capabilities and inspire future studies of those systems.

This paper is structured as follows. In Sect.~\ref{sc:data}, we describe the datasets used in this analysis. Section~\ref{sc:method} presents the methods, the adaptation of MCMF to the \Euclid data, and current limitations. In Sect.~\ref{sc:results}, we present results of running MCMF on \Euclid data from redshifts and richness in Sect.~\ref{sc:redshifts}, the forecast on cluster confirmation in Sect.~\ref{sc:forecast} and individual discussions on high-redshift clusters in Sect.~\ref{sc:indivclusters}. The paper closes with conclusions and an outlook on the future \Euclid DR1 dataset in Sect.~\ref{sc:concOutl}.
Throughout this paper we adopt a flat $\Lambda$CDM cosmology with $\Omega_\mathrm{m}=0.3$ and $H_0=70$\,km\,s$^{-1}$\,Mpc$^{-1}$.

\section{\label{sc:data} Data sets}
The survey fields contained in the \Euclid Q1 data release overlap with a huge number of optical and ICM-based cluster samples because of their location within the SDSS or DES optical surveys and their relative proximity to the northern or southern ecliptic poles. Therefore, a selection to a subset of cluster catalogues is needed.
In this work, we limit the set of cluster samples to the two largest X-ray catalogues to date, as well as the largest SZE-selected cluster catalogue and the SZE catalogue with the highest number of high-$z$ SZE-detected clusters in Q1. A further advantage of those surveys is the similarity in their optical follow-up, all using a red sequence-based technique together with imaging and photometric data from the Dark Energy Camera \citep[DECam; ][]{Flaugher15} to confirm clusters from an ICM-based candidate list.

\subsection{The \Euclid Q1 dataset}
The \Euclid Q1 release encompasses three fields, corresponding to the Euclid Deep Field North (EDF-N), the Euclid Deep Field South (EDF-S) and the Euclid Deep Field Fornax (EDF-F). The three fields have sizes of approximately 23 deg$^2$, 28 deg$^2$ and 12 deg$^2$ and the depth of the fields corresponds to that expected for the Euclid Wide Survey. As EDF-N does not overlap with deep SZE surveys or with the eROSITA eRASS1 survey and also differs in ground-based external photometry, we decided not to consider EDF-N for this study.
The remaining two fields share the same $g,r,i,z$-band external imaging data from the DECam with most of the contributing observations coming from DES. This makes both fields have very similar imaging quality in \Euclid- and ground-based datasets.
The data products and an overview of the Q1 data release can be found in \cite{Q1overview} and more details on the data processing of the VIS, NISP instruments, and the external (EXT) datasets can be found in dedicated papers from \citet{Q1VIS}, \citet{Q1NIRprocessing}, and \citet{Q1MER}, respectively. We therefore limit the discussion of the Q1 dataset to the most relevant details and refer the interested reader to the aforementioned publications. The \Euclid VIS instrument \citep{EucVisInst} observes the sky in a single band, \IE, ranging from 530 to 920 nm with an imaging resolution of 0.16\arcsec\ FWHM. The VIS data are used for optical investigation, as well as for star-galaxy separation and flagging. The NISP instrument \citep{EucNispInst} provides imaging in the three near infrared bands \YE, \JE, and \HE. The $5\sigma$ imaging depths are 26.0, 23.9, 24.1 and 24.0 for \IE, \YE, \JE, and \HE respectively \citep{Q1MER}.

\subsection{X-ray surveys}
The X-ray emission is proportional to the square of the electron density in the ICM and is therefore well suited to detect galaxy clusters without suffering projection effects caused by non-collapsed structures. The observed X-ray flux further depends on the inverse square of the luminosity distance, which causes the X-ray selected cluster samples to have a selection function
that is strongly redshift dependent. For this reason, the X-ray selected cluster samples preferentially probe the low-redshift and low-mass regime and therefore can be seen as the low-redshift anchor for this study.

\subsubsection{RASS-MCMF}
The RASS-MCMF cluster catalogue \citep{RASSMCMF} covers $\sim$\num{25000} deg$^2$ of good extragalactic sky. The ROSAT \citep{Truemper93} X-ray-based catalogue contains \num{8449} clusters and has a purity of 90\%. Catalogue creation closely follows the work on the predecessor catalogue MARD-Y3 \citep{MARDY3} and uses the MCMF cluster confirmation tool described in \citet{2018MNRAS.474.3324K,MARDY3} together with DESI Legacy Survey DR10 optical imaging and photometric data \citep{Legacysurveys19}. In contrast to other X-ray surveys, RASS-MCMF does not impose any cuts on apparent size of the X-ray emission. It uses advanced statistical techniques to control and measure sample purity using optical cluster richness. 

\subsubsection{eROSITA eRASS1}
The eROSITA \citep{eROSITA} X-ray telescope is the successor of the ROSAT telescope with increased image resolution and sensitivity. The telescope observed the whole sky more than four times \citep{Coutinho} until it was placed to safe mode due to the Russian invasion of Ukraine on 28th February 2022. The German eROSITA Consortium recently publicly released data and source catalogues of the first all-sky scan \citep{eRASS1Marloni} that covers the entire western Galactic hemisphere. Part of this release is the eROSITA eRASS1 catalogue of galaxies and groups covering an area of \num{12791} deg$^2$ and containing \num{12247} optically confirmed clusters and groups with expected purity of 86\%. The sample is constructed by imposing a cut on X-ray extent likelihood of $L_\mathrm{ext}>3$ and requiring the presence of an overdensity of red sequence galaxies in addition to the X-ray source detection. 

\subsection{SZE surveys}
The strength of the SZE signal depends on the line-of-sight integral of the electron pressure of the ICM and is similar to X-ray observations less prone to projection effects than other cluster probes such as optical or weak gravitational lensing-based studies. The SZE effect is redshift independent, which makes samples such as those used in this work have an approximately flat selection function with redshift. However, the SZE surveys used in this work suffer from atmospheric effects and the signal from the primary CMB, causing them to lose sensitivity at low redshift when angular sizes of clusters approach sizes of the primary CMB. The SZE surveys used in this work preferentially probe the redshift range of $0.2<z<2$. The SZE surveys therefore play a key role in this work in testing the performance of the \Euclid dataset at high redshift.

\subsubsection{SPT-SZ MCMF}
The SPT-SZ survey was conducted between 2007 and 2011 with the SPT-SZ camera on the South Pole Telescope \citep[SPT;][]{SPT} covering $\sim$\num{2500} deg$^2$. The first SPT-SZ cluster catalogue covering the full footprint contained 677 cluster candidates and 516 confirmed clusters above a signal-to-noise (SNR) threshold 4.5 was presented in \citet{Bleem15}. The SPT-SZ MCMF catalogue \citep{SPTSZMCMF} is based on this dataset and is the result of systematic optical and infrared follow-up of all candidates with $\mathrm{SNR}>4$ using homogeneous imaging data from DES and WISE \citep{Wright_2010,UnWISE}. Homogeneous follow-up, together with confirmation through the MCMF confirmation tool, allowed for an increased cluster sample of 811 clusters with a measured purity of 91\%. The completeness with respect to pure SZE selection is estimated to be 95\%. For homogeneity reasons, the SPT-SZ MCMF catalogue does not incorporate individual pointed observations, such as those in the infrared regime from the \textit{Spitzer} telescope \citep{Spitzer}.

\subsubsection{ACT-DR5 MCMF}
The ACT-DR5 MCMF catalogue \citep{klein2024actdr5} is based on data from the fifth data release \citep{ACTDR5data} from the Atacama Cosmology Telescope \citep{ACTtel}. Using homogeneous optical follow-up from the DESI Legacy Survey DR10 \citep{Legacysurveys19} and the most recent WISE imaging data together with the MCMF cluster confirmation tool over a total area of $\sim$\num{13000} deg$^2$. Compared to its predecessor catalogue \citep{ACTDR5Hilton}, the SZE detection
run were modified to ensure an unbiased selection at a signal-to-noise ratio of $\mathrm{SNR}=4$ and the more complete and systematic confirmation resulted in a significantly increased cluster sample with a higher completeness at high redshift. The total sample has \num{6247} clusters with an estimated purity of 89.3\%.

\section{\label{sc:method} Methods}
The MCMF cluster confirmation tool was developed in \cite{2018MNRAS.474.3324K} and refined in \cite{MARDY3} to confirm ICM-selected cluster candidates from large surveys. It was used to construct X-ray-selected and optically confirmed galaxy cluster catalogues from the ROSAT survey \citep{2018MNRAS.474.3324K,MARDY3,RASSMCMF} and eROSITA \citep{eFEDSKlein}. It was further used to create SZE-based galaxy cluster catalogues from \textit{Planck} \citep{2023MNRAS.525...24H}, SPT \citep{SPTSZMCMF,2024OJAp....7E..13B}, and ACT \citep{klein2024actdr5}. In addition to other work, MCMF-based catalogues were successfully used to derive cosmological constraints using weak gravitational lensing calibrated cluster number counts analysis \citep{2023MNRAS.522.1601C,Bocquet24b,2024arXiv240913556V,mazoun2024interactingdarksectorethos}.

In this work, we adopt the MCMF tool to the Q1 dataset to predict the ability of \Euclid data to confirm ICM-selected cluster candidates as real clusters. 
A further aim of this work is to probe the quality of the published ICM-selected samples at high redshift. We measure key properties of the clusters, namely the cluster redshift and the cluster richness $\lambda$ for the samples listed above of known clusters and compare them to published measurements to get an estimate of the performance of the \Euclid data on those properties. In a second step, we run MCMF towards random lines of sight and compare those measurements with our previous work using other imaging surveys, such as the DESI Legacy Survey \citep[LSDR10][]{Legacysurveys19} and unWISE \citep{UnWISE}. In Fig.~\ref{fig:acthighzd} we show three colour composite images of the cluster ACT-CL\,J0350.0$-$4819 using LSDR10, unWISE, and \Euclid\ to provide the reader a visual comparison of the data previously used for cluster confirmation and what is now available through \Euclid\ imaging data.

\subsection{MCMF}
The MCMF algorithm incorporates a red sequence technique \citep{gladders00,Rykoff14} that simultaneously uses redshift- and magnitude-dependent colour filters in the optical and infrared bands such as $g-r$, $r-i$, $i-z$, $z-w1$ in the case of LSDR10. Galaxies are further weighted according to their projected distance to the ICM-based centre and are only considered within a projected distance of $r_{500}$. Here $r_{500}$ is the radius within which the mean density is 500 times the critical density of the Universe at a given redshift, and is derived from the expected mass $M_{500}$ of a cluster candidate given redshift and ICM-based mass observable. 

\begin{figure}
\begin{center}
\includegraphics[width=1\linewidth]{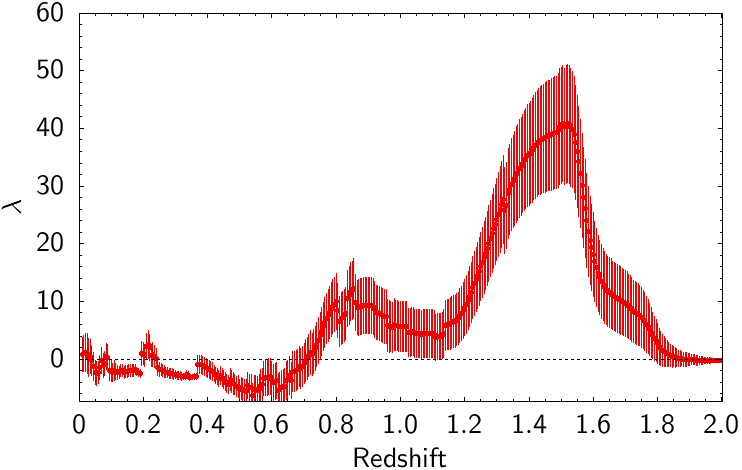}
\caption{$\lambda-z$ plot along the line of sight of ACT-CL\,J0350.0$-$4819, the ICM-selected cluster
with the highest redshift measured in the sample.}
\label{fig:actlamz}
\end{center}
\end{figure}

MCMF calculates richness in fine redshift steps along the cluster candidate line of sight and searches for peaks in the distribution, corresponding to possible optical counter parts to the ICM selected cluster. An example of such a measurement is shown in Fig.~\ref{fig:actlamz} towards the high-$z$ galaxy cluster ACT-CL\,J0350.0$-$4819 at $z\sim1.46$. We refer the interested reader to previous MCMF related papers \citep[e.g.][]{2018MNRAS.474.3324K,MARDY3} for a more detailed description on the richness and redshift measurements of MCMF. In the following subsection, we restrict the discussion to the modifications performed with respect to our latest work \citep{klein2024actdr5}.

\subsubsection{\label{sec:MCMFoEUC} MCMF on \Euclid}
For this work we reuse the red sequence models in the $g-r$, $r-i$, $i-z$ colours from previous MCMF work, which used the same DECam imaging data as those used in Q1 \citep{MARDY3,RASSMCMF}. We cross-match the LSDR10 and Q1 datasets and adopt small ($<0.025$ mag) shifts to observed galaxy colours to ensure overall colour compatibility. 
The small area of Q1 does not include a sufficient number of galaxy clusters with known spectroscopic redshifts that would allow direct self-calibration at high redshifts.  
We therefore calibrate the \Euclid-related red sequence colours ($z-\YE$, $\YE-\JE$, $\JE-\HE$) using the COSMOS photo-$z$ catalogue \citep{2022ApJS..258...11W}. We apply a cut in the specific star-formation rate ($<10^{-12}$ yr$^{-1}$) and stellar mass ($>10^{11}$ M$_\odot$) to preselect passive galaxies in the relevant mass range and further add 363 spectroscopically confirmed cluster members in the Q1 fields to increase the sample of galaxies at $z<0.1$. 
To check if the calibration the red sequence colours are reasonable, we run MCMF in different configurations omitting some of the bands. We find that redshifts remain consistent even in the case of dropping $g-r$ and $r-i$ colours, hence putting more weight to the remaining \Euclid-based bands.

While testing redshift recovery predominantly probes the agreement between predicted and observed red sequence colour, testing the recovered richness probes the model of the scatter of galaxies around the red sequence and with that the photometry.
In doing so, we found that the richness is not reasonably well recovered when different bands are omitted in the richness estimate. In fact, adding ground-based $g$- and $r$-band observations causes the richness to be significantly underestimated at low redshifts ($z<0.5$). 
Exploring the reasons for this, we identified at least two issues that impact the richness at low redshifts. We first identified that the scatter of the red sequence colour is significantly higher than that found in LSDR10, for the same DECam imaging data. In this context, we tested multiple types of photometry available in Q1 \citep{Q1MER} and their scatter in colour with respect to LSDR10 colours. We found that the template fitting-based photometry (\textit{TEMPLFIT}) provides the lowest scatter for red sequence-like galaxies compared to the other photometry provided in the Q1 catalogue, but still yielding significantly higher scatter than LSDR10 photometry. We account for this fact by increasing the intrinsic scatter of the red sequence by a factor of two in our models. 

The second issue is related to flux measurements for bright but compact galaxies between 17 and 21 magnitudes in the \YE band. The current \Euclid processing is not optimised for those sources, and cores of those galaxies are masked and not properly processed. As a result $\sim20\%$ of the galaxies in that magnitude range around clusters show huge offsets in at least one of the colours at levels of $\sim0.5$ magnitudes in \textit{TEMPLFIT}-based measurements. More details on known issues with the Q1 data can be found in the Q1 Explanatory Supplement.\footnote{https://euclid.esac.esa.int/dr/q1/expsup} These offsets correspond to several sigma in terms of total expected scatter of the red sequence galaxies. Galaxies affected by this issue are heavily down-weighted in the richness estimate as they appear not to be consistent with the red sequence at the cluster redshift, and hence the richness is biased low.
Exploring other types of photometry, we found that the Q1 aperture photometry is affected by the same issue, while the S\'ersic profile-based photometry appears significantly less affected. 
We therefore use S\'ersic profile-based photometry measurements to replace \textit{TEMPLFIT}-based colours in $g,r,i,z$-bands in cases where colours deviate more than 0.2 magnitudes from each other. 
While this reduces the impact of this issue to the richness estimate, it does not completely resolve the problem.
As a way to mitigate this problem, we therefore decided to use the full set of colours for redshift estimates while ignoring the $g-r$ and $r-i$ colours for the richness estimate to avoid too strong biases in the low-redshift regime. We remind the reader that the main focus of the paper is predominantly in the $z>1$ regime, where ground-based imaging data lose their constraining power and the photometry of bright sources in the $g,r,i,z$-bands is not of relevance.

MCMF typically performs model fits to the peaks found in the $\lambda$ vs. $z$ measurement (see Fig.~\ref{fig:actlamz}) towards each candidate position. The used peak models are generally based on a large set of clusters with spectroscopic redshifts. This allows for improved redshift and richness estimates by accounting for the expected shape of a peaks, as well as it provides the ability to de-blend structures in redshift space. They further provide an in-build calibration to spectroscopic redshifts ensuring unbiased cluster redshifts.

As the Q1 fields are too small to construct reliable peak models, we decided to quote redshifts and richnesses based on the peaks identified in the $\lambda$ vs. $z$ measurements.
The addition of the \Euclid NIR bands ensures better behaved peak shapes at high redshift, reducing the need for performing model fits to the data. De-blending in redshift space is only relevant for a small percentage of the clusters sample and will not be of high importance for this work either. The first main data release (DR1) will incorporate improvements on the issues found in Q1 and provide sufficient area to perform self-calibration of red sequence and peak models like it was done in previous work.

\begin{figure}
\begin{center}
\includegraphics[width=1\linewidth]{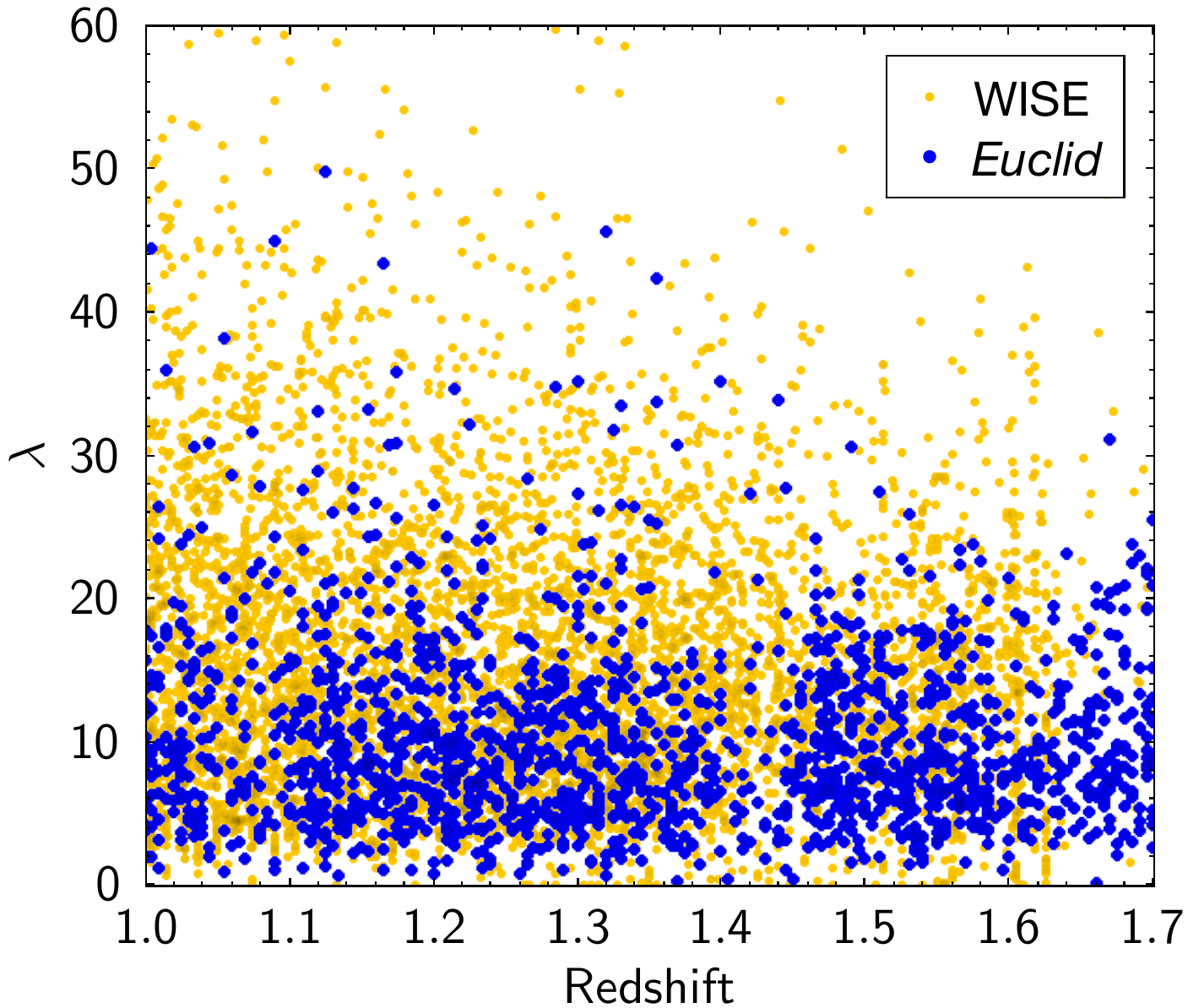}
\caption{Comparison of the richness distribution along random lines of sight as a function of redshift using either WISE-based or \Euclid-based MCMF measurements. No strong redshift evolution is visible for \Euclid-based MCMF measurements. The lower scatter for \Euclid-based MCMF suggests lower noise and hence improved separation of real clusters from contaminants.}
\label{fig:randoms}
\end{center}
\end{figure}

\subsection{\label{sc:rand}MCMF towards random lines of sight}

MCMF employs the redshift and richness distributions derived from random lines of sight, alongside those of the actual ICM-selected candidates, to distinguish genuine clusters from false detections. The redshifts and richnesses of these random lines of sight are assumed to represent the background population of false positives within the ICM-selected candidate list. To best approximate this contaminating population, it is necessary to account for the fact that MCMF uses an ICM-based mass proxy to estimate richness, and that the survey data themselves contain real ICM-selected clusters. Consequently, the appropriate configuration of these so-called randoms depends on the survey from which the ICM-selected candidates originate.

Given the focus of this work on assessing {\it Euclid\/}’s performance at high redshift, we generated randoms following the characteristics of the SZE-based ACT-DR5 MCMF catalogue. Specifically, we constructed a set of random positions that exhibits a signal-to-noise distribution similar to that expected from noise fluctuations in ACT-DR5, excluding regions around confirmed ACT-DR5 clusters.

This approach should be applicable to other deep SZE-based surveys as well. For even deeper SZE surveys, we expect the richness measured along random lines of sight to decrease. This is due to the fact that the expected mass at fixed SZE signal-to-noise becomes smaller, which in turn reduces the aperture size ($r_{500}$) used by MCMF to compute richness. Additionally, more randoms that by chance fall near real clusters are excluded, as these clusters now exceed the detection threshold of the given survey. In total, we ran MCMF on approximately \num{5000} random positions within Q1. In Fig.~\ref{fig:randoms} we show the redshift and richness measurements of the randoms in direct comparison to similar WISE-based MCMF randoms for the $z>1$ range. The \Euclid-MCMF measurements show lower richnesses, suggesting that the noise due to non-collapsed or low-mass systems is significantly smaller when compared to WISE-based measurements. The precise impact on cluster confirmation will be discussed in Sect.~\ref{sc:forecast}.

\begin{figure*}
\begin{center}
\includegraphics[height=0.445\linewidth]{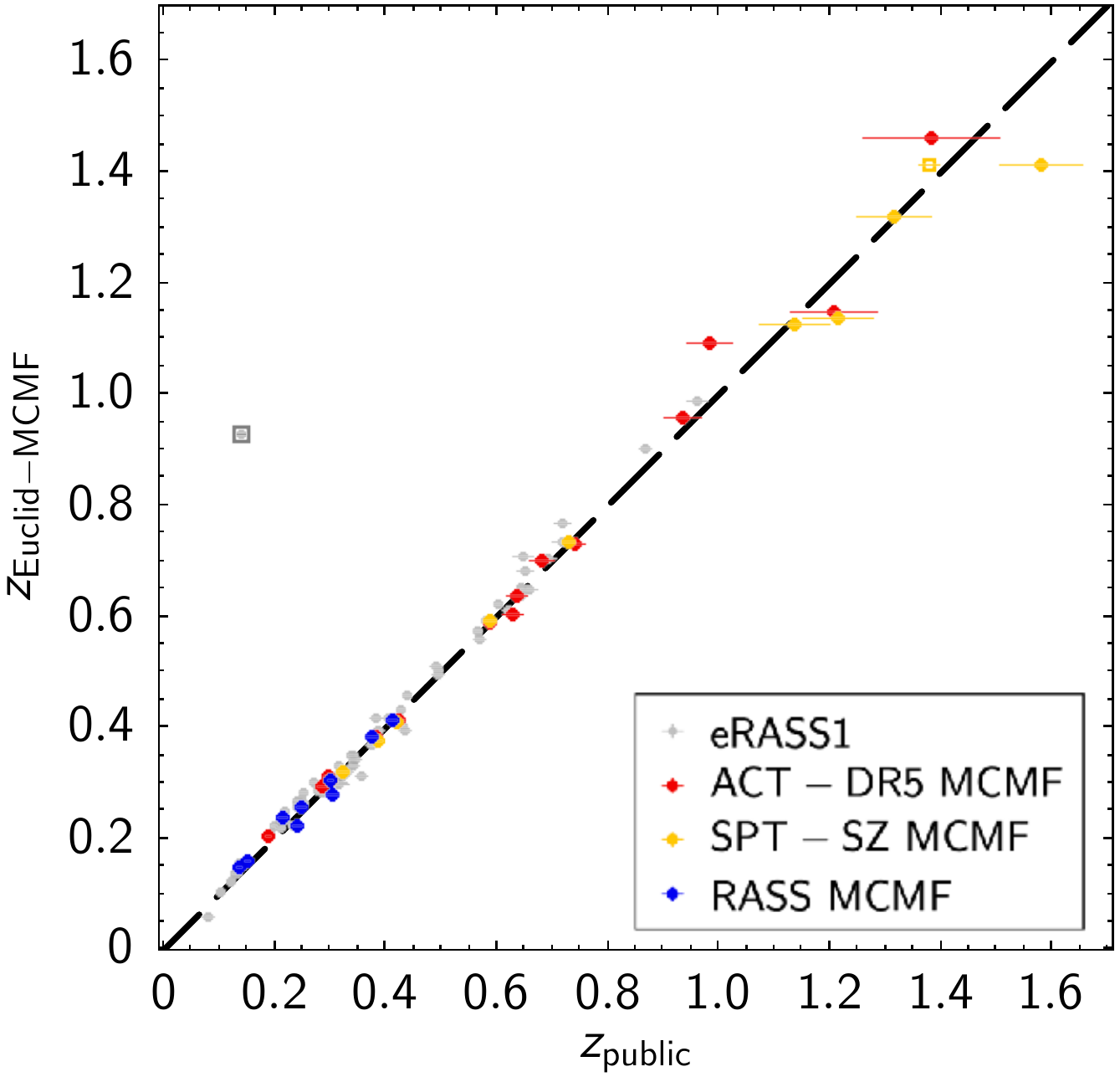}
\includegraphics[height=0.45\linewidth]{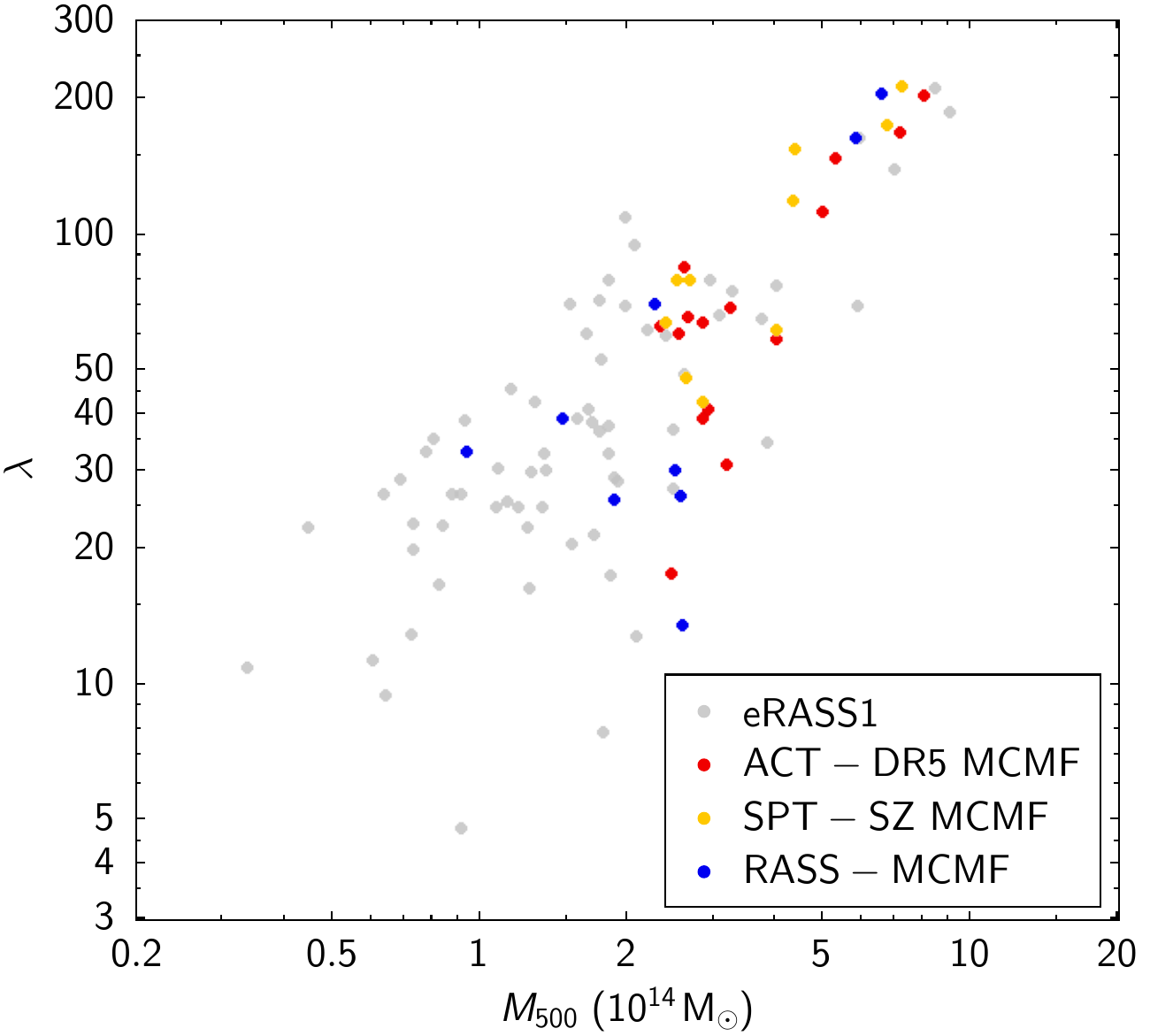}
\caption{\textit{Left:} Recovery of published cluster redshifts with \Euclid-MCMF. Redshifts of 1eRASS\,J040527.1$-$490347 (grey square) and SPT-CL\,J0421$-$4845 (yellow square) are discussed in greater detail in Sect.~\ref{sc:indivclusters}. \textit{Right:} \Euclid-MCMF richness ($\lambda$) versus ICM-based mass estimates ($M_{500}$).}
\label{fig:redshiftRichness}
\end{center}
\end{figure*}

\section{\label{sc:results} Results}
\subsection{\label{sc:redshifts} Redshift and richness}

MCMF was run without fitting peak profiles to the richness-versus-redshift distributions along the candidate lines of sight. This results in a slight reduction in performance and introduces the possibility of small biases between the true and inferred redshifts. Furthermore, it prevents de-blending of multiple structures along the line of sight. The performance of MCMF on \Euclid data is therefore expected to improve further as the survey progresses and sufficient area for self-calibration becomes available. Using the combined sample of ACT-DR5 MCMF, SPT-SZ MCMF, RASS-MCMF, and eRASS1, we correct for a 3\% bias in the mean \Euclid-MCMF photo-$z$, assuming that the ensemble of catalogues exhibits negligible bias with respect to the true cluster redshift.

The left panel of Fig.~\ref{fig:redshiftRichness} compares the \Euclid-MCMF photo-$z$ with published measurements. As shown in the plot, we generally find good agreement between the published and \Euclid-MCMF results. We identify 1eRASS\,J040527.1$-$490347, marked as a grey square in Fig.~\ref{fig:redshiftRichness}, as having a significantly different redshift. This cluster, together with the five clusters with published redshifts greater than one, is discussed in more detail in Sect.~\ref{sc:indivclusters}. The cluster with the highest published redshift, SPT-CL\,J0421$-$4845, shows a $\sim2.1\sigma$ deviation from the \Euclid-MCMF estimate. This deviation is likely due to a bright point source affecting the WISE-based redshift estimate reported in the SPT-SZ MCMF catalogue. This cluster also has deep HST and \textit{Spitzer} imaging data, with an improved redshift estimate of $z=1.38\pm0.02$ \citep{Strazzulo19}, and is shown as a yellow square in Fig.~\ref{fig:redshiftRichness}. As stated in Sect.\ref{sc:data}, all datasets make use of DECam imaging data. This may cause some correlation in the photo-z estimates between those catalogues and the \Euclid\-based measurements, especially at low redshift. At $z>1$ the infrared datasets gain more relevance, reducing any remaining correlation between measurements.

In the right panel of Fig.~\ref{fig:redshiftRichness}, we show the \Euclid-MCMF richness as a function of ICM-based masses from the four cluster samples analysed in this work. To ensure approximate comparability between masses from different surveys, we use the overlap of the full cluster catalogues (not limited to the Q1 field) to derive simple scaling factors. As a reference, we adopt the calibrated $M_{500}$ estimates from ACT-DR5 MCMF \citep{klein2024actdr5}, as described in more detail in the original ACT-DR5 catalogue paper \citep{ACTDR5Hilton}. As seen in Fig.~\ref{fig:redshiftRichness}, the \Euclid-MCMF richness correlates strongly with ICM-based mass estimates.

We visually inspected the three lowest-richness clusters in the sample. One cluster is affected by masking due to a bright star, another may be impacted by an X-ray point source, and the third is at a very low redshift ($z=0.13$), where current issues with the Q1 photometric data may compromise the richness measurement.

\subsection{\label{sc:forecast} Forecasting confirmation performance at high redshift}

As the number of real ICM-selected cluster candidates is small, we must base the forecast of confirmation performance solely on the properties of the measurements along random lines of sight. As shown in Fig.~\ref{fig:randoms}, the richness distribution along random lines of sight is approximately redshift-independent for the \Euclid-based MCMF.

\begin{figure*}
\begin{center}
\includegraphics[width=0.38\linewidth]{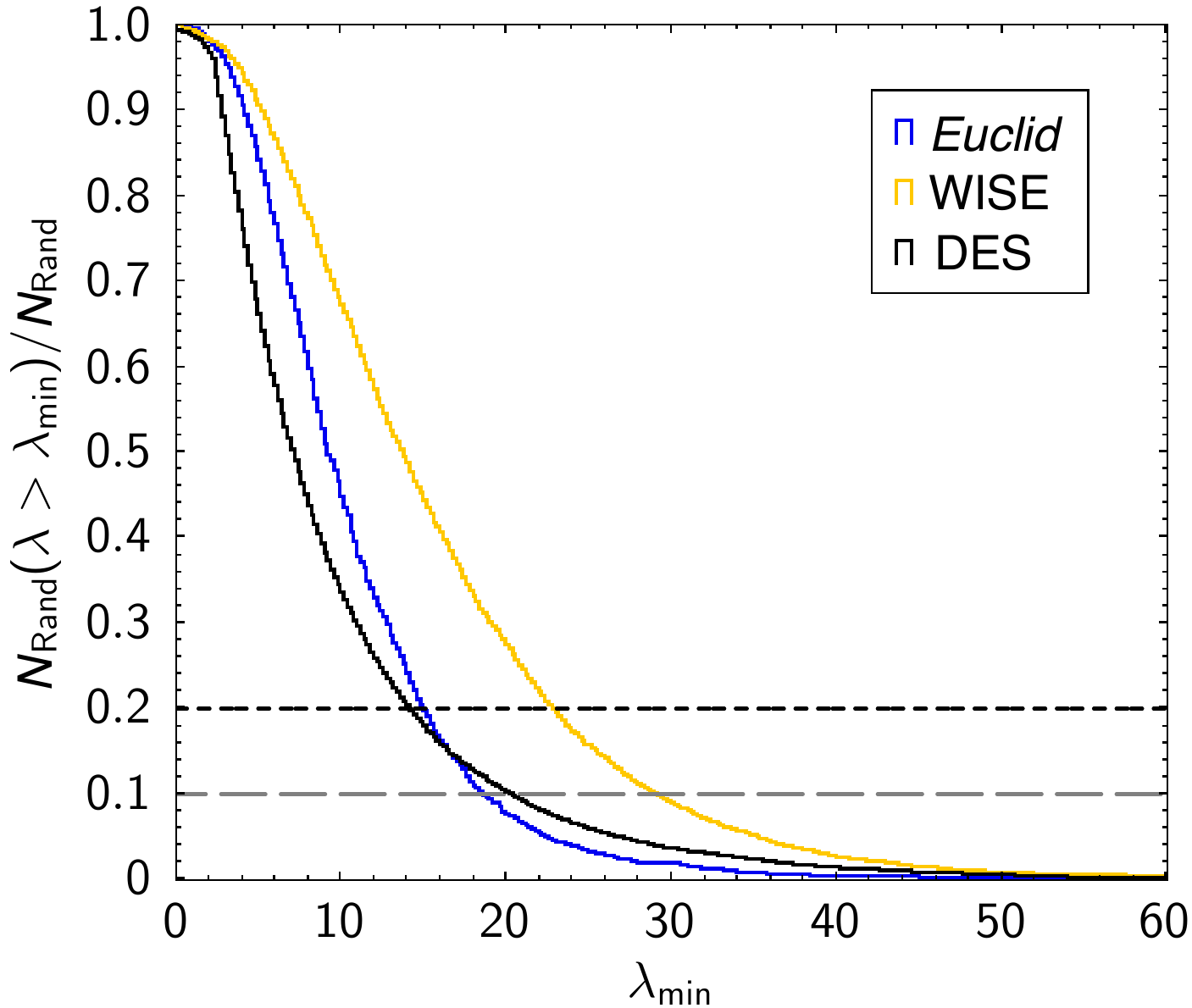}
\includegraphics[width=0.61\linewidth]{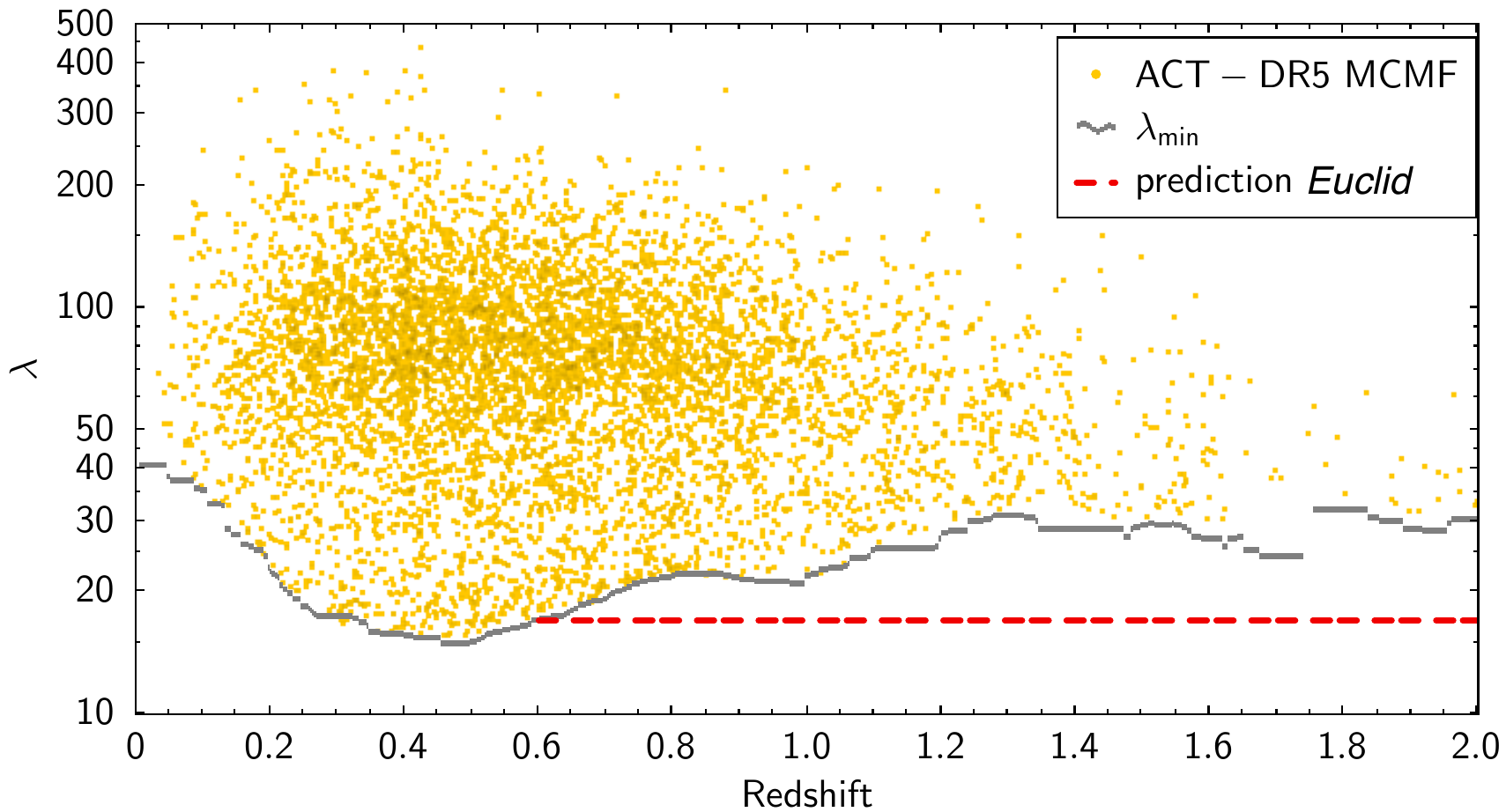}
\caption{\textit{Left:} The richness distribution along random lines of sights and above $z>1$ from \Euclid-MCMF is compared to WISE-based MCMF above $z>1$ as well as DES-based MCMF at $z<0.6$. The richness distributions of \Euclid-MCMF follows closely that of low-$z$ DES measurement and significantly improves over WISE. \textit{Right:} Distribution of ACT-DR5 MCMF clusters in richness vs. redshift. The properties of \Euclid-MCMF measurements along random lines of sight suggest that future \Euclid-based confirmation will allow similar performance at $z>1$ than current day DES-based confirmation at $z<0.6$.}
\label{fig:forecast}
\end{center}
\end{figure*}

From Sect.~\ref{sc:redshifts} and Fig.~\ref{fig:redshiftRichness}, we found no evidence that the richness is biased or degraded with respect to the original richness estimates from ACT-DR5 MCMF. This allows us to directly compare the richness distributions along random lines of sight from the original ACT-DR5 MCMF work \citep{klein2024actdr5} with those derived in this study. These richness distributions are shown in the left-hand panel of Fig.~\ref{fig:forecast}, illustrating the fraction of randoms exceeding a certain richness threshold ($\lambda_{\mathrm{min}}$). Shown are the \Euclid Q1 measurements at $z>1$, compared to the WISE-MCMF version and those derived from DES- or LSDR10-like MCMF measurements at $z<0.6$.

The WISE-MCMF measurements were previously used to confirm most ACT-DR5 MCMF clusters at $z>1$, where the ground-based DES/LSDR10 photometry was not sufficiently deep. Comparing WISE-MCMF with \Euclid-MCMF, it is evident that the richness distribution for \Euclid-MCMF declines much more steeply. This allows one to exclude 80\% of chance superpositions at richness values of approximately $\sim15$, compared to $\sim25$ for WISE-MCMF. Moreover, the richness distribution is in close agreement with that of ground-based data at $z<0.6$, where the imaging depth of those surveys is sufficient for cluster confirmation.

The right-hand panel of Fig.~\ref{fig:forecast} shows the redshift and richness distribution of the ACT-DR5 MCMF catalogue, including the redshift-dependent richness cut used to construct the sample and ensure a constant level of purity as a function of redshift. This selection exhibits several features related to the limitations of the different surveys used to construct the sample. The upturn in $\lambda_{\mathrm{min}}$ at low redshift is due to the SZE selection function and the reduced ability to detect very extended clusters at low $z$. The minimum of $\lambda_{\mathrm{min}}$ occurs around $z\approx0.5$, where the bulk of the real ACT-DR5 clusters are located, followed by an increase towards $z\approx1.2$ due to limitations of the optical survey. At higher redshifts ($z\geq1.1$), WISE-based confirmation becomes more effective than ground-based data, causing the richness threshold to remain approximately constant.

The richness distribution along random lines of sight for \Euclid-MCMF being approximately similar to that of lower-redshift optical surveys suggests that, for ACT-DR5-like surveys, the redshift-dependent richness cut could remain at the level currently only achieved at redshifts of $z\approx0.6$. While current SZE surveys such as ACT-DR5 MCMF are not strongly impacted by incompleteness due to their relatively high mass thresholds and the small fraction of high-redshift clusters, future samples from surveys such as SPT-3G \citep{SPT3G}, SO \citep{SObs}, and CMB-S4 \citep{CMBS4white} will benefit significantly from the enhanced high-redshift cluster confirmation capabilities offered by \Euclid data.

\subsection{\label{sc:indivclusters} Discussion on individual high-$z$ clusters}
Compared to ground-based data, \Euclid offers uniquely high imaging resolution along with simultaneously deep \YE, \JE, \HE imaging. These capabilities become especially important for distant clusters, where the rest-frame optical bands are redshifted into the infrared regime, and where imaging resolution becomes crucial for resolving individual cluster member galaxies. In the following subsection, we therefore discuss all high-redshift ($z>1$) clusters in our sample of ICM-selected clusters.

\subsubsection{1eRASS\,J040527.1$-$490347}
\begin{figure}
\begin{center}
\includegraphics[width=1\linewidth]{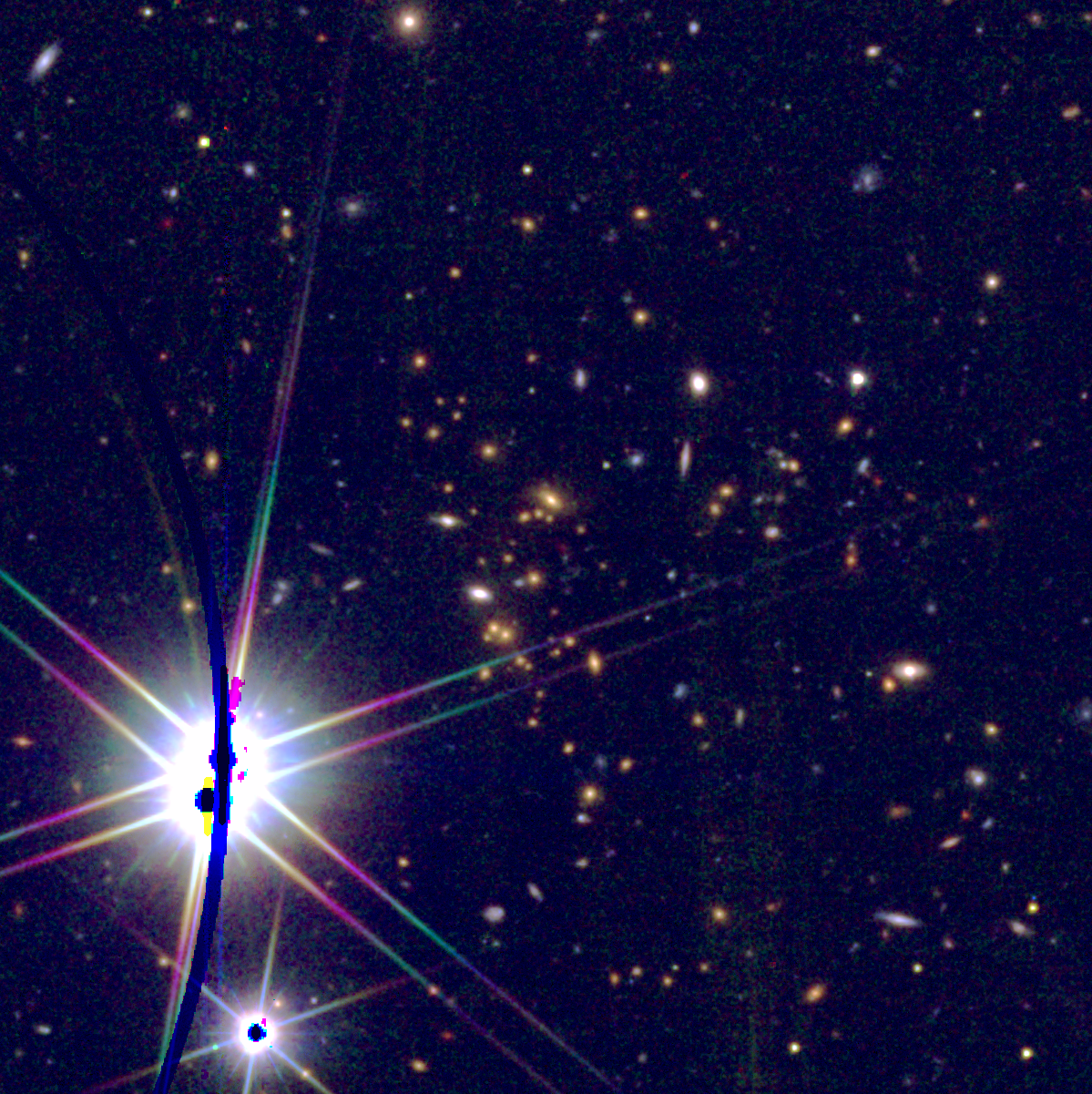}
\caption{\Euclid \IE, \YE, \HE-band colour composite image of the central $2\arcmin \times 2\arcmin$ region of 1eRASS\,J040527.1$-$490347 ($z=1.1$). While the bright stars in the south-east corner may have impacted previous confirmation and may act as additional X-ray emitters, a massive cluster is clearly visible in the \Euclid imaging data.}
\label{fig:eRASShighz}
\end{center}
\end{figure}

The \Euclid-MCMF measurement of 1eRASS\,J040527.1$-$490347 reveals a rich cluster ($\lambda=76$) at $z=1.1$. In Fig.~\ref{fig:eRASShighz}, we show the \IE, \YE, \HE-band colour composite of the central region around the cluster. According to the redshifts given in the eRASS1 cluster catalogue \citep{ERASS1}, no eRASS1 cluster in the Q1 field is expected to exceed $z=1$. The redshift of $z=0.254\pm0.015$ listed for 1eRASS\,J040527.1$-$490347 in \citet{ERASS1} is inconsistent with the redshift derived using \Euclid photometry. 

The most likely reason for this mismatch is the presence of bright point sources in the vicinity of the cluster, combined with the high redshift of the system. These factors likely caused the cluster galaxies to be either masked out or too faint to be detected in the ground-based imaging data alone.

\subsubsection{SPT-CL\,J0354$-$4815}

SPT-CL\,J0354$-$4815 is the cluster with the lowest SZE significance among all $z>1$ clusters in Q1. With a cluster richness of $\lambda\approx75$, it is well detected in the \Euclid data. The redshift of $z=1.13$ is in good agreement with the value listed in the SPT-SZ MCMF catalogue ($1.14\pm0.06$). This redshift is independently confirmed through a nearby cross-match with the optical cluster catalogue by \citet{Wen24}, who report $z=1.16$ for this cluster.

\subsubsection{ACT-CL\,J0353.0$-$4818, SPT-CL\,J0353$-$4818}

\begin{figure}
\begin{center}
\includegraphics[width=0.99\linewidth]{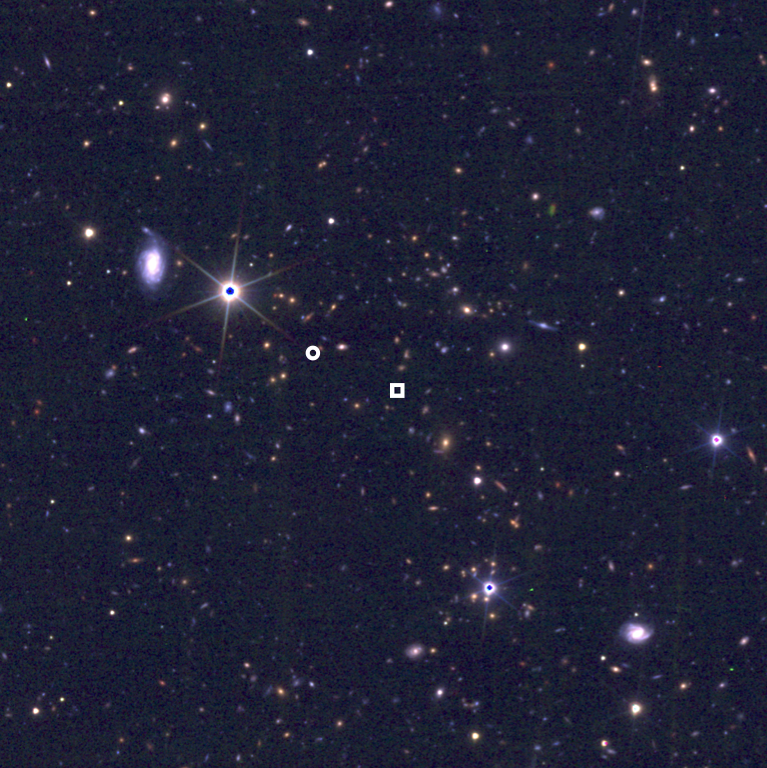}
\caption{\Euclid \IE, \YE, \HE-band colour composite image of the a $3\arcmin \times 3\arcmin$ region around ACT-CL\,J0353.0$-$4818 (SPT-CL\,J0353$-$4818). The ACT-DR5 position is marked by a white square, while the SPT-SZ position is marked by a white circle. The offset between both SZE-based positions is close to the typical offset found between SPT-SZ and ACT-DR5 clusters.}
\label{fig:SPT0353-48}
\end{center}
\end{figure}

The cluster ACT-CL\,J0353.0$-$4818, also known as SPT-CL\,J0353$-$4818, is robustly detected in both SZE surveys. The \Euclid-based MCMF analysis yields a redshift of $z=1.13$ when using SPT priors, and a highly consistent value of $z=1.14$ when using ACT-DR5 priors. These estimates align well with those from ACT-DR5 MCMF ($z=1.21\pm0.08$), SPT-SZ MCMF ($z=1.22\pm0.06$), and the value reported by \citet{Wen24} ($z=1.165$) using independent optical cluster finding. 
Earlier versions of the ACT-DR5 and SPT-SZ catalogues \citep{ACTDR5Hilton, Bleem15} reported lower redshifts for this system: $z=0.93\pm0.03$ and $z=0.96$, respectively. Despite being a well-detected SZE cluster, the appearance of this system in \Euclid imaging is less prominent. As shown in Fig.~\ref{fig:SPT0353-48}, red galaxies are sparsely distributed over an extended region elongated along the north-east to south-west axis. The BCG lies approximately 17\arcsec\ from the ACT-DR5 position and around 36\arcsec\ from the SPT position. This optical morphology suggests that the cluster is not a relaxed system.

\subsubsection{SPT-CL\,J0353$-$5043}
SPT-CL\,J0353$-$5043 is a well-detected SPT cluster with an SPT signal-to-noise ratio of $\xi=5.35$ and a redshift of $z=1.315\pm0.067$ according to SPT-SZ MCMF. The \Euclid\ MCMF estimate yields $z=1.32$, fully consistent with the published value. The SZE position lies 11\arcsec\ from a large passive galaxy, which likely corresponds to the BCG of the system.

\begin{figure*}
\begin{center}
\includegraphics[width=0.45\linewidth]{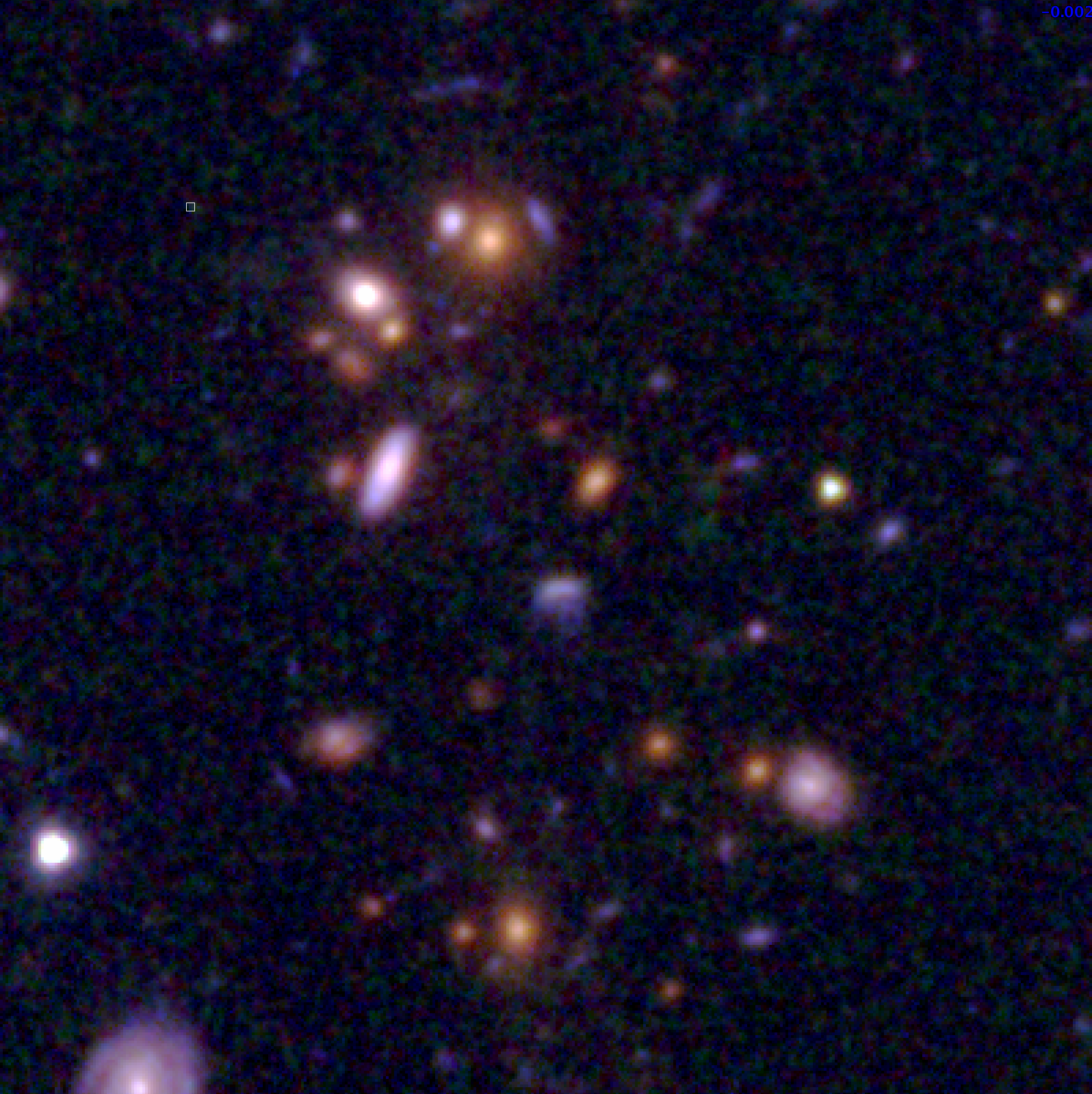}
\includegraphics[width=0.45\linewidth]{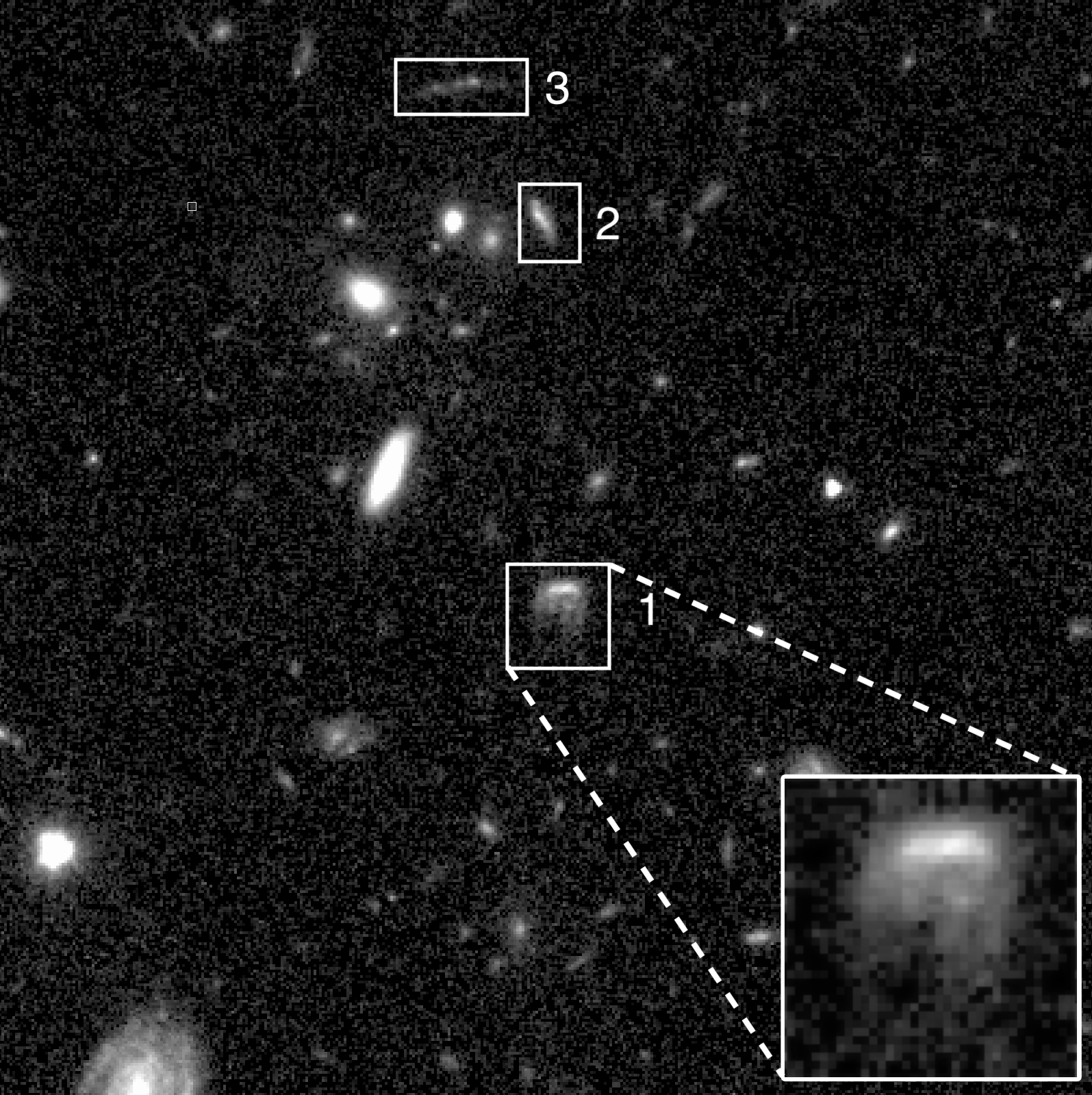}
\caption{\textit{Left:} \Euclid \IE, \YE, \HE-band colour composite image of the central $40\arcsec \times 40\arcsec$ region of SPT-CL\,J0353$-$5043. \textit{Right:} Grey scale image of the same region in \Euclid \IE. Boxes mark the highest redshift jellyfish candidate to date (1) and two strongly lensed galaxies (2,3).}
\label{fig:SPT0353-50}
\end{center}
\end{figure*}

The left panel of Fig.~\ref{fig:SPT0353-50} shows a colour composite image of the central region of SPT-CL\,J0353$-$5043 using the \Euclid\ VIS \IE and NIR \YE and \HE bands. The BCG is visible in the northern half of the image, while the field also includes a second bright galaxy in the southern region. The field of view corresponds to approximately 350\,kpc. The right panel of Fig.~\ref{fig:SPT0353-50} shows the \IE band alone in greyscale, marking three interesting galaxies. Galaxies labelled 2 and 3 are likely arcs from strongly lensed background galaxies. Additional tentative lensing features are visible near the BCG and the southern elliptical galaxy. 
A comparison of the colour and \IE-band images reveals that cluster members, particularly the BCG, appear faint in the \IE band relative to the NISP bands. This behaviour results from the \IE band probing the rest-frame UV regime ($240$–$390$\,nm) at the cluster redshift, which is sensitive to recent star formation. The faintness of the BCG in this band suggests a lack of significant ongoing star formation and supports the picture that BCG's in massive clusters at redshifts $z\sim1.3$ are predominantly passive galaxies. For cluster lenses at these redshifts this yields a significantly increased contrast in \IE between lensed star-forming galaxies and their lenses dominated by passive galaxies. We refer the interested reader to \cite{Q1stronglensclusters} for a dedicated search for strong-lensing galaxy clusters in Q1.

In contrast, a jellyfish-shaped blue galaxy (EUCL\,J035330.86$-$504347.6) is visible near the centre of the image. This galaxy, marked with number 1 in the right-hand panel of Fig.~\ref{fig:SPT0353-50}, has a photometric redshift of $z=1.34\pm0.03$ from the \Euclid\ photo-$z$ pipeline, which is highly consistent with the cluster redshift. 
Jellyfish galaxies are formed through interactions between the intracluster medium (ICM) and the interstellar medium of infalling galaxies, which experience strong ram-pressure stripping \citep{Jellyfish_rampressure_Gunn,EnviroenmenLateType_Boselli}. Due to the high infall velocity, the galaxy’s gas is stripped and trails behind, forming a tail in the direction opposite to its motion. Under favourbale conditions, star formation can happen within this stripped tail, making such galaxies bright in optical and UV bands, with the star-forming disc and the tail giving a jellyfish appearance \citep{Jellyfish_Owen,Jellyfish_cortese,Jellyfish_Owers,Jellyfsihebeling,Jellyfish_poggianti,Jellyfish_koshy}.
This scenario is consistent with the morphology observed in Fig.~\ref{fig:SPT0353-50}, where the galaxy appears to be moving towards the BCG and cluster centre, with the stripped gas, hosting ongoing star formation, trailing in the opposite direction. Its shape, colour, photometric redshift, and the direction of the stripped component, strongly suggest that EUCL\,J035330.86$-$504347.6 is indeed a jellyfish galaxy at $z\approx1.32$.
Spectroscopic redshift and kinematic information of the disc and tails are required to confirm the ram-pressure stripping nature of the galaxy. If validated, this galaxy could represent one of the highest-redshift jellyfish galaxies found in an ICM-selected cluster to date.

\begin{figure*}
\begin{center}
\includegraphics[width=1\linewidth]{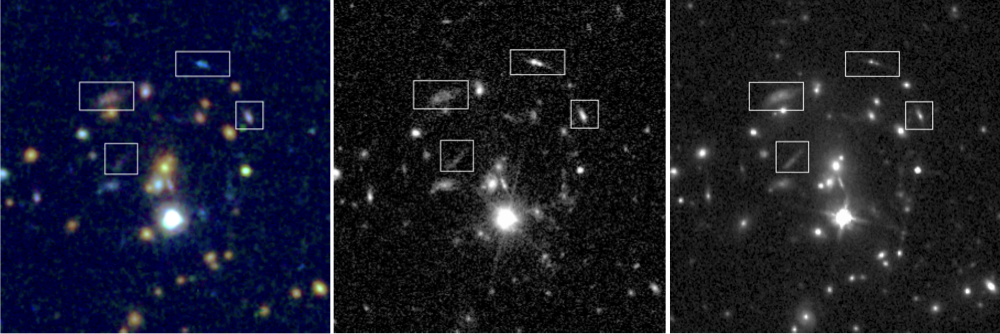}
\caption{Central $30\arcsec \times 30\arcsec$ region of SPT-CL\,J0421$-$4845. The \Euclid \IE, \YE, \HE colour composite image shown left, the \Euclid \IE image in the central, and the HST F140W image is shown in the right panel. Marked are four candidates of strongly lensed galaxies.}
\label{fig:SPT0421}
\end{center}
\end{figure*}

\subsubsection{SPT-CL\,J0421$-$4845}

SPT-CL\,J0421$-$4845 belongs to a subset of five $z>1.4$ clusters that were studied in greater detail by \citet{Strazzulo19} using \textit{Hubble} imaging data. The original photometric redshift of $z=1.42\pm0.09$ from the SPT-SZ cluster catalogue \citep{Bleem15} was refined to $z=1.38\pm0.02$ by \citet{Strazzulo19}, placing the cluster slightly below the original redshift selection threshold of $z=1.4$ for their sample. Using our MCMF adaptation to \Euclid\ data, we find $z=1.41$, in good agreement with previous estimates.
In contrast, the redshift reported in the SPT-SZ MCMF catalogue, $z=1.58\pm0.08$, based on Legacy Survey DR10 and WISE data, is $\sim2\sigma$ higher than the values found by other studies. A likely explanation is a bright AGN located approximately 5\arcsec\ south of the BCG, which severely affects the WISE photometry in the cluster core. The red appearance of this source in WISE bands likely biases the redshift estimate toward higher values.

The central region of SPT-CL\,J0421$-$4845, shown in Fig.~\ref{fig:SPT0421}, reveals several galaxies that appear to be strongly lensed by the cluster potential. In addition to the \Euclid\ \IE, \YE, \HE colour composite (left panel) and the \IE image (central panel), we include an HST F140W image with a total exposure time of 2400\,s. Four galaxies, marked with white rectangles, exhibit tangential alignment relative to the cluster centre and have colours consistent with background galaxies, making them strong lensing candidates.
Despite the availability of HST imaging, no dedicated strong or weak lensing analysis has yet been conducted for this cluster. The combination of existing HST and new \Euclid\ imaging makes SPT-CL\,J0421$-$4845 a compelling target for cross-instrumental lensing studies.

\subsubsection{ACT-CL\,J0350.0$-$4819}

ACT-CL\,J0350.0$-$4819, shown in Fig.~\ref{fig:acthighzd}, is the highest-redshift cluster identified by our MCMF analysis of \Euclid\ data, with a photometric redshift of $z=1.46$, consistent with the ACT-DR5 MCMF value of $z=1.38$ within the expected uncertainties. Given this high redshift, ground-based data are nearing their limit in redshift determination. In the case of the WISE-based redshift estimate, two blue galaxies near the cluster core are blended with the core of the cluster, complicating accurate photometric measurements for this cluster. 
This highlights the advantage of the high-resolution infrared imaging capabilities of \Euclid compared to current large-scale surveys.

\begin{figure}
\begin{center}
\centering
\includegraphics[width=1\linewidth]{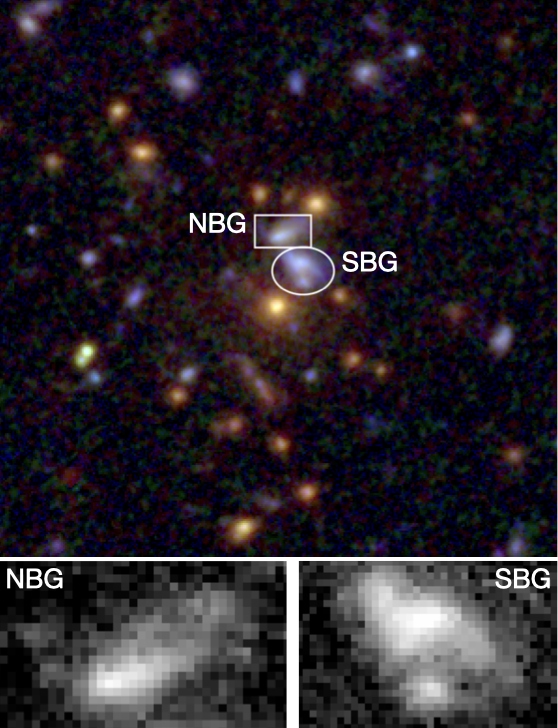}
\caption{ \Euclid \IE, \YE, \HE colour composite of a  $\ang{;0.5;}\times\ang{;0.5;}$ region centred on the cluster centre of ACT-CL\,J0350.0$-$4819 at the top, zoomed \Euclid \IE-band cutouts of the northern blue galaxy (NBG) and the southern blue galaxy (SBG) are shown below.}
\label{fig:ACT0350zoom}
\end{center}
\end{figure}

In Fig.~\ref{fig:ACT0350zoom}, we take advantage of {\it Euclid\/}’s excellent image resolution and present a $30\arcsec \times 30\arcsec$ image of the cluster core. Near the central passive galaxy, we mark two blue galaxies. Visual inspection reveals similar colours for both galaxies, but photometric redshifts from \Euclid\ differ significantly. The northern blue galaxy (NBG; EUCL\,J035000.73$-$481943.8) has a photometric redshift of $z=1.55$, while the southern blue galaxy (SBG; EUCL\,J035000.64$-$481945.4) has $z=0.05$.
This difference in redshifts might be explained by degeneracies in colour space between low- and high-redshift galaxies visbile as redshift outliers at low photometric redshifts \citep{EUCQ1photoz}.
Their irregular morphologies and the small angular separation between both galaxies in fact suggests that both galaxies are likely interacting with each other. The small separation to the cluster BCG together with redshift consistency of the NBG with the cluster redshift further supports a scenario where both star-forming galaxies are associated with ACT-CL\,J0350.0$-$4819. Additional data, such as spectroscopic redshifts, are needed to confirm the nature of this system and its relation to the cluster.

\section{\label{sc:concOutl}Conclusions and Outlook}

In this work, we investigate the capabilities of \Euclid\ in the follow-up and confirmation of ICM-selected galaxy clusters, using data from the \Euclid\ quick release (Q1). We measure \Euclid-based redshifts and richnesses for clusters drawn from four of the largest ICM-selected samples to date: RASS-MCMF, SPT-SZ MCMF, ACT-DR5 MCMF, and eROSITA eRASS1.

Our adaptation of the MCMF algorithm to \Euclid\ Q1 data demonstrates reasonable performance in both redshift and richness estimation and is expected to perform even better with the upcoming DR1 dataset.
While redshift measurements perform well across the full range, we encounter challenges in richness estimation for low-redshift clusters. These are partially mitigated by excluding $g$- and $r$-band colours from the richness calculation.

The redshift estimates, particularly at high redshift ($z>1$), where ground-based optical surveys typically lose sensitivity, show excellent agreement with previously published values. While this confirms the general quality of the samples used in this work, we also identify one case where the superior imaging quality of \Euclid revises a published MCMF-based redshift. In the case of SPT-CL\,J0421$-$4845, our \Euclid-based MCMF analysis confirms the redshift of $z\sim1.4$ reported by \citet{Bleem15} and \citet{Strazzulo19} based on HST and \textit{Spitzer}, while revealing that the WISE-based MCMF redshift in the SPT-SZ catalogue \citep{SPTSZMCMF} is biased high by $\sim2\sigma$, due to AGN contamination near the cluster core.

Richness estimates derived from the \Euclid-based MCMF show generally high correlation with ICM-based mass estimates. However, the performance remains suboptimal due to challenges in Q1 photometry and the current inability to self-calibrate red sequence models within the small Q1 dataset. These issues are expected to be resolved in future releases. Nonetheless, even with the current Q1 dataset, {\it Euclid\/}’s high-redshift performance, particularly in the NISP bands, is already outstanding. 

By applying MCMF to random lines of sight, we show that the noise properties of \Euclid-based richnesses at $z>1$ are comparable to those achieved by modern ground-based surveys at $z<0.6$. This marks a significant improvement over WISE-based cluster confirmation currently driving high-$z$ cluster confirmation and demonstrates {\it Euclid\/}’s potential to identify the majority of high-redshift ICM-selected clusters from future surveys.

We further leverage {\it Euclid\/}’s deep, high-resolution imaging data to study the $z>1$ cluster subsample in more detail. Notably, we revise the redshift of one eRASS1 cluster 1eRASS\,J040527.1$-$490347 from $z=0.254\pm0.015$ to $z=1.1$, with the discrepancy likely due to masking by bright stars in the vicinity of the cluster. All five SZE-selected clusters at $z>1$ are reconfirmed to be at high redshift. In SPT-CL\,J0353$-$5043 and SPT-CL\,J0421$-$4845, we find evidence for strong lensing features, and in ACT-CL\,J0350.0$-$4819, we identify two massive, star-forming galaxies near the BCG. Spectroscopic follow-up is needed to confirm their cluster membership and investigate their possible interaction or merger with the BCG.

Finally, we report the discovery of the highest-redshift jellyfish galaxy in an ICM-selected cluster known to date. The galaxy EUCL\,J035330.86$-$504347.6, located less than 20\arcsec\ from the BCG of SPT-CL\,J0353$-$5043 at $z=1.32$, has a photometric redshift consistent with the cluster. The \Euclid\  \IE-band imaging, probing the rest-frame UV ($240$–$390$\,nm), shows clear evidence of ram-pressure stripping. The extended stellar emission points away from the cluster centre, consistent with jellyfish morphology. If confirmed with spectroscopic redshift and kinematic information of the disc and tail, this would push the redshift frontier for jellyfish galaxies from $z=0.8$ \citep{HighzJelly_durret} to $z=1.32$, thereby setting constraints on the role of environment-driven star formation quenching in high-redshift galaxies. The combination of broad VIS-band coverage and high imaging quality in \Euclid\ is likely well suited for identifying jellyfish galaxies in high-redshift clusters, making this discovery potentially the first of many in the upcoming wide survey.

The increased area covered by the Euclid Wide Survey, together with recent and upcoming deep SZE surveys like those based on SPT-3G \citep{SEDFS,SPT3G} will result in a significant increase of the number of ICM-selected galaxy clusters at high redshifts. Besides being valuable probes for cosmological and astrophysical studies, the overlap with \Euclid\-selected galaxy clusters will further allow to investigate systematics related with either surveys. The availability of the SZE-based mass proxy out to high redshift can further be used to test and improve optical mass proxies such as the cluster richness estimated by MCMF or by the \texttt{RICH-CL} function \citep{RICHclmethod} within \Euclid\ galaxy cluster workflow \citep{Q1-SP050}.

%
%

\begin{acknowledgements}
This work has made use of the Euclid Q1 data from the {\it Euclid} mission of the European Space Agency (ESA), 2025,
\url{[https://doi.org/10.57780/esa-2853f3b|https://doi.org/10.57780/esa-2853f3b]}.

\AckEC  
\end{acknowledgements}

%
%

\bibliography{Euclid,  myRefs} 

%

  

%

\label{LastPage}
\end{document}